\documentclass[preprint,12pt]{elsarticle}

\usepackage{multirow}

\usepackage[T1]{fontenc} 
\usepackage{lmodern} 
\usepackage{graphicx} 
\usepackage{amsmath, amssymb, amsfonts} 
\usepackage{amsthm} 
\usepackage{mathrsfs} 
\usepackage{anyfontsize}
\usepackage[title]{appendix} 
\usepackage{xcolor} 
\usepackage{textcomp} 
\usepackage{manyfoot} 
\usepackage{booktabs} 
\usepackage{algpseudocode} 
\usepackage{mathtools} 
\usepackage{float} 
\usepackage{bm} 

\usepackage{threeparttable} 
\usepackage{subcaption} 
\usepackage[section]{placeins} 
\usepackage{array} 
\usepackage{soul} 
\usepackage{xcolor}
\usepackage{enumitem}

\usepackage{relsize}
\newcommand{\cross}{\times}

\journal{Journal of Computational Science}

\begin{document}

\begin{frontmatter}



\title{SABR-Informed Multitask Gaussian Process: A Synthetic-to-Real Framework for Implied Volatility Surface Construction}


\author[inst1]{Jirong Zhuang}

\affiliation[inst1]{organization={Department of Mathematics, University of Macau},
            addressline={Avenida da Universidade Taipa}, 
            postcode={999078}, 
            state={Macau},
            country={China}}

\author[inst1]{Xuan Wu*}

\begin{abstract}

This study introduces a SABR-informed multitask Gaussian process for constructing implied volatility surfaces from sparse option quotes. We treat a dense synthetic dataset generated by a calibrated SABR model as the source task and market option quotes as the target task. Within the multitask Gaussian process framework, we learn cross-task dependence via task embeddings with hierarchical regularization, enabling adaptive transfer of structural information. On Heston ground truth across ten market regimes and in a case study with SPX options, the model achieves lower error than the single-task Gaussian process and SABR at near-term maturities and remains competitive at long-term maturities, while satisfying standard no-arbitrage conditions. The framework combines the theory-driven structure with nonparametric Bayesian regression and yields reliable implied volatility surfaces for risk management.

\end{abstract}


\end{frontmatter}



\section{Introduction}

 Constructing the implied volatility surface (IVS) from sparse option quotes is a standard task in practice. Trading desks use the surface to support volatility quoting and the day-to-day risk management of vanilla option books across strikes and maturities. Structuring teams and model developers use it as an input to the calibration of downstream pricing models. The task is particularly challenging when market liquidity is limited, for example around the market open or for less liquid option products, and in illiquid regions such as far-from-the-money strikes and long maturities where observations are sparse and unevenly distributed. These operational requirements motivate surface construction methods that remain stable under data sparsity while producing smooth and no-arbitrage extrapolations in the wings.

Structural models, such as the SABR model \citep{Hagan2002}, are popular because they capture volatility smiles and skews with few parameters.  These models assume that the underlying asset and its volatility follow specific stochastic differential equations. While these models provide reasonable interpolations, their predetermined functional forms may lack the flexibility to fully reflect market complexities. In contrast, data-driven approaches such as the Gaussian process (GP) \citep{refWilliams2006}, a nonparametric Bayesian method that can be used in IVS construction \citep{Spiegeleer2018,roberts2021probabilistic}, offer more flexibility. However, this adaptability requires sufficient data. With sparse observations, they can be prone to overfitting and may violate no-arbitrage conditions.

In order to address the challenges in data-driven approaches, recent works discuss how to incorporate prior financial knowledge into machine learning methods. The literature \citep{cousin2022gaussian,chataigner2021beyond,Ackerer2020,zheng2021incorporating,gonon2024operator,hoshisashi2023no} presents methods that impose constraints derived from financial theory within a neural network or a Gaussian process. In contrast, \citet{chen2023teaching} use transfer learning methods for option pricing. A neural network is trained on synthetic data generated from the Black-Scholes model, and it provides a good initialization for subsequent fine-tuning on empirical data.

Inspired by the transfer learning approach of \citet{chen2023teaching}, we propose a SABR-informed multitask Gaussian process (SABR-MTGP). The key idea is to treat the SABR-generated IVS as an information source (a `source' task) and market observations as the primary target task. Then, a multitask learning framework \citep{bonilla2007multi, alvarez2012kernels} is used where the relationship between the theoretical structure (SABR) and empirical observations is learned adaptively during a joint optimization process.  In our experiments, the proposed SABR-MTGP stabilizes data-sparse wings (far-from-the-money strikes) relative to a single-task GP, while correcting systematic misspecification that can arise from purely SABR models. The implementation consists of two main stages. First, we generate a synthetic dataset using the calibrated SABR model. This dataset embodies smile and term-structure patterns characteristic of SABR dynamics. Second, we train a multitask Gaussian process (MTGP) model simultaneously on these dense synthetic data and the sparse market observations. The MTGP framework learns the correlation between tasks through the shared covariance component. We use task embeddings with hierarchical regularization to appropriately balance how much structural guidance from SABR is used when constructing the IVS from real market data. SABR-MTGP improves IVS reconstruction from sparse quotes, but it is not a replacement for SABR-based dynamic hedging, or OTC derivatives valuation. Our scope is IVS construction from sparse observations. Extending to hedging and OTC derivatives valuation requires market-consistent dynamics (e.g., local or stochastic local volatility) and Greek validation. These are left to future work.

\subsection*{Related Literature}

Implied volatility surface construction can be organized into three complementary approaches. First, structural models specify dynamics that reproduce smiles and term structures with interpretable parameters, including SABR \citep{Hagan2002}, Heston \citep{heston1993closed}, Variance Gamma \citep{madan1998variance} and the double-exponential model \citep{kou2002jump}. More recently, rough volatility models \citep{bayer2016pricing,gatheral2022volatility} emphasize fractional regularity in volatility paths. Second, parametric surface parameterizations impose shape and no-arbitrage structure, e.g., SVI \citep{gatheral2004parsimonious}, SSVI \citep{gatheral2014arbitrage}, and spline-based families \citep{corlay2013b}. Third, machine learning approaches, including the Gaussian process \citep{Spiegeleer2018,roberts2021probabilistic,chataigner2021beyond} and neural networks \citep{Ackerer2020,zheng2021incorporating,gonon2024operator,hoshisashi2023no,ning2023arbitrage,cont2023simulation} provide data-driven alternatives. 

Relative to transfer learning approaches for options (\citet{chen2023teaching}), our contribution is to implement the transfer learning in the Gaussian process via a task kernel with learnable embeddings and a hierarchical prior that adapts cross-task correlation.

Besides implied volatility surface construction, Gaussian processes have found increasing application across finance and economics due to their flexibility and ability to quantify uncertainty. A significant body of work focuses on derivative pricing, valuation, and hedging. GP models have been used for constructing financial term structures \citep{Cousin2016}, accelerating pricing and hedging calculations for various options \citep{Spiegeleer2018, Ludkovski2018, goudenege2020machine, li2022effect, hocht2024pricing}, modeling derivative portfolios for CVA computation \citep{Crepey2020}, pricing complex insurance products like variable annuities \citep{goudenege2021gaussian}, and approximating Greeks for hedging \citep{ludkovski2021}. Beyond direct pricing, GP models serve as flexible tools for calibrating implied and local volatility surfaces \citep{roberts2021probabilistic,chataigner2021beyond} or measuring portfolio tail risk \citep{risk2018sequential}. Furthermore, GP models are employed in broader financial and economic time series analysis, including imputing missing financial data \citep{de2020gaussian}, forecasting real estate prices \citep{xu2023gaussian}, modeling inflation dynamics \citep{clark2024forecasting}, understanding determinants of carbon market prices \citep{salvagnin2024investigating}, and developing nonparametric vector autoregressions for macroeconomic analysis \citep{hauzenberger2025gaussian}.

The remainder of the paper is organized as follows: Section \ref{sec:background} introduces basic concepts of the implied volatility surface. Section \ref{sec:methodology} provides the details of the proposed methods. In Section \ref{sec:experiments}, we describe the setup of the numerical experiments and the results are presented in Section \ref{sec:results}. In Section \ref{sec:spx_application}, we demonstrate the application of the proposed methods using real market data.

\section{Background: Implied Volatility Surface} \label{sec:background}

The Black-Scholes model \citep{black1973pricing} is a basic framework for European option pricing under specific assumptions: the underlying asset price follows a geometric Brownian motion, and the asset's volatility is constant. In practice, this constant volatility assumption is inconsistent with observed option prices, which typically exhibit patterns such as volatility smiles and skews. To reconcile the Black-Scholes formula with market prices, the concept of Implied Volatility (IV) is introduced. For an observed market price of call options $C_{mkt}(K, \tau)$ with time-to-maturity $\tau$ and strike price $K$, the corresponding IV $\sigma_{mkt}(K,\tau)$ is the unique positive value that makes the following equation hold:
    \begin{align}
    C_{mkt}(K, \tau) = C_{BS}(K, \tau, \sigma = \sigma_{mkt}(K,\tau)\,;\theta_{BS}),
    \end{align}
    where $C_{BS}$ is the Black-Scholes formula, and $\theta_{BS}$ is the collection of remaining parameters in the Black-Scholes formula.
    Computing $\sigma_{mkt}(K,\tau)$ requires inverting the Black-Scholes formula using iterative root-finding algorithms (such as Newton-Raphson). When $\sigma_{mkt}(K,\tau)$ is shown as a function of both variables, the resulting three-dimensional surface is the Implied Volatility Surface (IVS). The patterns of IVS reflect market participants' risk preferences and expectations about future price dynamics beyond the standard Black-Scholes assumptions. Accurate IVS construction is important for derivative pricing, hedging strategies, and risk management across option portfolios.

\section{SABR-Informed Multitask Gaussian Process} \label{sec:methodology}

In this section, we describe our approach for constructing the implied volatility surface by combining SABR structure with GP flexibility.

\subsection{Problem Formulation}
The objective of this work is to construct an accurate Implied Volatility Surface (IVS). The IVS is a function of the option strike price $K$ and its time-to-maturity $\tau$. We can represent the input features as a two-dimensional vector $\mathbf{x} = [K, \tau]^\intercal$. The task is to learn a regression model $f: \mathbb{R}^2 \to \mathbb{R}^+$ that maps these input features to the corresponding implied volatility.

Suppose that we have $N$ observed locations $\mathbf{X}_{\mathcal{T}}=[\mathbf{x}_{\mathcal{T},1},\dots,\mathbf{x}_{\mathcal{T},N}]^\intercal$, $\mathbf{x}_{\mathcal{T},i}=[K_i, \tau_i]^\intercal$, and $\mathbf{y}_\mathcal{T}=[y_{\mathcal{T},1},\dots,y_{\mathcal{T},N}]^\intercal$ contains the corresponding observed market implied volatilities, $y_{\mathcal{T},i} = \sigma_{\rm mkt}(K_{i},\tau_i)$. We denote the market-observed option dataset as $\mathcal{D}_\mathcal{T} = \{(\mathbf{X}_\mathcal{T}, \mathbf{y}_\mathcal{T})\}$. A significant challenge in practice is that this market data $\mathcal{D}_\mathcal{T}$ is often sparse, particularly for options with long maturities or strikes far from the current underlying price (deep out-of-the-money or in-the-money). This sparsity can make it difficult for purely data-driven models to construct reliable IVS.

To address this challenge, our methodology transfers information from a structural financial model, specifically the SABR model. The core idea is to generate a dense synthetic dataset, denoted $\mathcal{D}_\mathcal{S}$, using a calibrated SABR model. This synthetic dataset acts as a source of information to guide the learning process and prediction, especially in regions where market data is scarce. The subsequent sections will show how this synthetic data is generated and incorporated within a multitask learning framework.

\subsection{Synthetic Data Generation via SABR}\label{sec_SABR}

A key step in our work was creating a synthetic dataset that captures structural patterns of volatility surfaces. In this paper, we used the SABR model \citep{Hagan2002} for incorporating financial theory to guide the model. SABR is one of the most widely used parametric models in quantitative finance for modeling implied volatility smiles and skews. Hagan's asymptotic expansion provides a closed-form approximation for implied volatility under SABR, which is crucial for efficiently generating the dense synthetic dataset $\mathcal{D}_\mathcal{S}$.

The SABR model describes the dynamics of the forward price $F_t\coloneqq S_te^{(r-q)(T-t)}$ (where $S_0$ is the current price of the underlying asset, $r$ is the risk-free interest rate, $q$ is the dividend rate, $T$ is the maturity) and its instantaneous volatility $\alpha_t$ via stochastic differential equations:
\begin{align}
dF_t &= \alpha_t F_t^\beta dW_t^{(1)}, \\
d\alpha_t &= \nu \alpha_t dW_t^{(2)}, \\
dW_t^{(1)} dW_t^{(2)} &= \rho dt.
\end{align}
Here, $\beta \in [0,1]$ relates volatility to the price level, $\nu$ is the volatility-of-volatility, and $\rho$ is the correlation between two Brownian motions. Each parameter influences the IVS shape: $\alpha$ determines the volatility level, $\beta$ affects the backbone shape, $\rho$ controls the skew, and $\nu$ drives smile convexity.

\subsubsection*{SABR Model Asymptotic Expansion}

Practical applications often use Hagan's asymptotic expansion \citep{Hagan2002}, which provides closed-form approximations for implied volatility $\sigma_{\rm SABR}(K, \tau, F; \alpha, \beta, \rho, \nu)$.  
\begin{equation}
\sigma_{\rm SABR}(K,\tau,F)=\left\{
\begin{aligned}
  &I^0(K,\tau,F)\left[1+I^1(K,F)\tau \right] \quad &\text{if}\, K\neq F, \\
    &\frac{\alpha}{F^{1-\beta}}
    \left[
    1+\left(\frac{(1-\beta)^2\alpha^2}{24F^{2-2\beta}}+
    \frac{\rho\beta\alpha\nu}{4F^{1-\beta}}+
    \frac{2-3\rho^2}{24}\nu^2\right)\tau
    \right] \quad &\text{if}\, K= F.
\end{aligned}\right. \label{eq:sabr_detailed}
\end{equation}
where:
\begin{align*}
    I^0(K,\tau,F)=&\left( \frac{\alpha z(K,F)}{\chi(z)(F K)^{(1-\beta)/2}} \right)
        \left[
        1 + \frac{(1-\beta)^2}{24}\log^2 (F/K) +\frac{(1-\beta)^4}{1920}\log^4 (F/K)
        \right]^{-1},\\
        I^1(K,F)=&\frac{(1-\beta)^2\alpha^2}{24(FK)^{1-\beta}}+
        \frac{\rho\beta\nu\alpha}{4(FK)^{(1-\beta)/2}}+\frac{2-3\rho^2}{24}\nu^2,\\
        z(K,F)=&\frac{\nu}{\alpha}(FK)^{(1-\beta)/2}\log (F/K),\quad\chi(z)=\log\left[\frac{\sqrt{1-2\rho z+z^2}+z-\rho}{1-\rho}\right].
\end{align*}

Let $ \mathbb{T}_\text{mkt}=\{\tau_{(1)}, \dots, \tau_{(P)}\}$ be the set of unique maturities in the market data $\mathcal{D}_\mathcal{T}$, ordered $ \tau_{(1)} < \tau_{(2)} < \dots < \tau_{(P)} $. For each maturity slice $\tau_{(p)}$, we calibrated SABR parameters $(\alpha_{(p)}, \rho_{(p)}, \nu_{(p)})$ to match the observed market smile by minimizing the squared errors using the Nelder-Mead algorithm \citep{nelder1965simplex}:
\begin{equation}
\min_{\alpha_{(p)}, \rho_{(p)}, \nu_{(p)}} \sum_{i:\,\tau_i=\tau_{(p)}} \left( \sigma_{\rm SABR}(K_i, F_{\tau_{(p)}}, \tau_{(p)};\, \alpha_{(p)}, \beta,\rho_{(p)},\nu_{(p)}) - y_{\mathcal{T},i} \right)^2
\end{equation}
Here, $ F_{\tau_{(p)}}=S_0e^{(r-q)\tau_{(p)}}$ is the forward price at maturity $ \tau_{(p)} $. Calibrating all four SABR parameters ($\alpha, \beta, \rho, \nu$) simultaneously for each maturity slice can be challenging and numerically unstable, especially with sparse or noisy data.  Calibrating $\beta$ often requires richer data across strikes than might be available, particularly for longer maturities. In practice, fixing the backbone parameter $\beta$ is not restrictive for smile fitting. The remaining SABR parameters $(\alpha,\rho,\nu)$ provide sufficient flexibility to reproduce the observed smile with high accuracy (\citet{WOS:000382423100006}), while improving calibration stability under sparse data. We fixed $\beta=0.5$ throughout our experiments. Robustness with respect to the choice of $\beta$ is assessed in Section~\ref{sec:sensitivity} (Figure~\ref{fig:beta_robustness}). The parameters $\alpha_{(p)}$, $\rho_{(p)}$, and $\nu_{(p)}$ were optimized within realistic ranges: $0.01 < \alpha_{(p)} < 2.0$, $-0.99 < \rho_{(p)} < 0.0$, and $0.05 < \nu_{(p)} < 1.5$. This yielded calibrated parameters $ \{ (\alpha_{(p)}, \rho_{(p)}, \nu_{(p)}) \}_{p=1}^P $ for each observed maturity slice.

To generate the synthetic dataset, we needed SABR parameters for any desired maturity. We defined a dense grid of $M$ points $\mathbf{X}_\mathcal{S}=[\mathbf{x}_{\mathcal{S},1},\dots, \mathbf{x}_{\mathcal{S},M}]^\intercal$, where $\mathbf{x}_{\mathcal{S},j} = [\tilde{K}_j, \tilde{\tau}_j]^\intercal$, covering the region of interest.

For any grid maturity $\tilde{\tau}$, we determined the SABR parameters as follows:
\begin{itemize}
    \item If $\tilde{\tau}$ matches a calibrated maturity $\tau_{(p)}$, we used the calibrated parameters $(\alpha_{(p)}, \rho_{(p)}, \nu_{(p)})$.
    
    \item If $\tilde{\tau}$ is between two calibrated maturities, $\tau_{(p)} < \tilde{\tau} < \tau_{(p+1)}$, we used piecewise linear interpolation to get parameters $(\alpha_{\tilde\tau}, \rho_{\tilde\tau}, \nu_{\tilde\tau})$. For example, for $\alpha$:
    \begin{equation} \label{eq:interpolation}
    \alpha_{\tilde\tau} = \alpha_{(p)} + \frac{\tilde{\tau}-\tau_{(p)}}{\tau_{(p+1)}-\tau_{(p)}}(\alpha_{(p+1)}-\alpha_{(p)})
    \end{equation}
    The same applied for $\rho_{\tilde\tau}$ and $\nu_{\tilde\tau}$.
    
    \item For extrapolation beyond the calibrated range ($\tilde{\tau} < \tau_{(1)}$ or $\tilde{\tau} > \tau_{(P)}$), we used constant extrapolation (parameters from the nearest calibrated maturity).
\end{itemize}

With parameters for each grid point, we computed synthetic volatilities $$y_{\mathcal{S},j} = \sigma_{\rm SABR}(\tilde{K}_j, F_{\tilde{\tau}_j},\tilde{\tau}_j;\alpha_{\tilde{\tau}_j}, \beta, \rho_{{\tilde\tau}_j}, \nu_{{\tilde\tau}_j}).$$ This produced the synthetic source dataset $\mathcal{D}_\mathcal{S} = \{(\mathbf{X}_\mathcal{S}, \mathbf{y}_\mathcal{S})\}$, where $\mathbf{y}_\mathcal{S}=[y_{\mathcal{S},1},\dots,y_{\mathcal{S},M}]^\intercal$. In practice, Gaussian noise $ \epsilon_{\mathcal{S}} \sim \mathcal{N}(0, \sigma^2_{\text{syn}})$ is added to the synthetic volatilities to prevent the model from becoming overly dependent on the synthetic data.

\subsection{Multitask Gaussian Process Formulation}

To combine information from the dense synthetic data and sparse market observations, we used a Multitask Gaussian process (MTGP) framework. 

A Gaussian process (GP) is a Bayesian nonparametric approach that can be used in regression problems \citep{refWilliams2006}. A GP defines a distribution over functions $f:\mathbb{R}^d\to\mathbb{R}$ such that any finite set of function values has a joint Gaussian distribution. Standard GP models handle a single output, while MTGP models handle multiple related outputs by capturing correlations between tasks \citep{bonilla2007multi, alvarez2012kernels}. 

We designated the synthetic SABR data $\mathcal{D}_\mathcal{S}$ as the source task and the sparse market data $\mathcal{D}_\mathcal{T}$ as the target task. For each task, observed volatilities were modeled as noisy realizations of a latent function:
\begin{equation}
    y_\mathcal{S}(\mathbf{x})=f_\mathcal{S}(\mathbf x)+\varepsilon_\mathcal{S}, \quad \varepsilon_\mathcal{S}\sim\mathcal{N}(0,\sigma^2_\mathcal{S}),
    \quad y_\mathcal{T}(\mathbf{x})=f_\mathcal{T}(\mathbf x)+\varepsilon_\mathcal{T}, \quad \varepsilon_\mathcal{T}\sim\mathcal{N}(0,\sigma^2_\mathcal{T})
\end{equation}
The noise variances $ \sigma^2_\mathcal{S} $ and $ \sigma^2_\mathcal{T} $ reflect different uncertainty levels in the synthetic SABR approximations and market observations.

A key idea was decomposing each latent function to facilitate information sharing. We expressed each task's function as:
\begin{equation}
    f_{\mathcal{S}}(\mathbf{x})=g_\mathcal{S}(\mathbf x)+h_{\mathcal S}(\mathbf x),\quad f_{\mathcal{T}}(\mathbf{x})=g_\mathcal{T}(\mathbf x)+h_{\mathcal T}(\mathbf x)
\end{equation}
This decomposition included:
\begin{itemize}
    \item Task-specific components $g_\mathcal{S}$ and $g_\mathcal{T}$ that captured unique patterns in each task
    \item Shared components $h_\mathcal{S}$ and $h_\mathcal{T}$ that contributed to the construction of correlations between tasks
\end{itemize}
We placed independent GP priors on the task-specific components, using a common input covariance function $k$ (e.g., a Matérn 5/2 kernel) with task-specific variance scaling parameters:
\begin{equation}
    g_\mathcal{S}\sim\mathcal{GP}(0,\kappa_{\mathcal S}^2k), \quad g_\mathcal{T}\sim\mathcal{GP}(0,\kappa_{\mathcal T}^2k)
\end{equation}
The parameters $\kappa_{\mathcal S}^2$ and $\kappa_{\mathcal T}^2$ control the amount of task-specific variance.

For the shared components, we used learned task embeddings to infer the relationship. We modeled $h_\mathcal{S}$ and $h_\mathcal{T}$ as jointly Gaussian with a covariance structure:
\begin{equation}
    \text{Cov}(h_{\mathcal Z}(\mathbf x),h_{\mathcal Z^\prime}(\mathbf x^\prime))= \underbrace{\sigma^2_h \exp\left(-\frac{\| \mathbf{e}_\mathcal{Z}-\mathbf{e}_{\mathcal{Z}^\prime}\|^2}{l_h^2}\right)}_{
    \eqqcolon\,\hat{C}_{\mathcal{Z},\mathcal{Z}^\prime} 
    }k(\mathbf x,\mathbf x^\prime) \quad \text{for}\quad \mathcal{Z,Z^\prime}\in\{\mathcal{S,T}\}
\end{equation}
The term $\hat{C}_{\mathcal{Z},\mathcal{Z}^\prime}$ depends on task embeddings $\mathbf{e}_\mathcal{Z}, \mathbf{e}_{\mathcal{Z}^\prime} \in \mathbb{R}^{d^\prime}$ ($d^\prime$ is a hyperparameter that needs to be predetermined), a shared variance scaling parameter $\sigma^2_h$ and a length-scale parameter $l_h$. This allowed the model to learn the level of information sharing. Tasks with closer embeddings would have stronger correlation through the shared component. With components $g_\mathcal{S}, g_\mathcal{T}, h_\mathcal{S}, h_\mathcal{T}$, the overall covariance structure follows an Intrinsic Coregionalization Model (ICM) \citep{bonilla2007multi}:
\begin{equation}
    \text{Cov}(f_\mathcal{Z}(\mathbf{x}), f_{\mathcal{Z}^\prime}(\mathbf{x}^\prime)) = \underbrace{\left( \hat{C}_{\mathcal{Z},\mathcal{Z}^\prime} + \kappa^2_\mathcal{Z}\delta_{\mathcal{Z}\mathcal{Z}^\prime} \right)}_{\eqqcolon\, C_{\mathcal{Z},\mathcal{Z}^\prime}} k(\mathbf{x}, \mathbf{x}^\prime)\quad \text{for}\quad \mathcal{Z,Z^\prime}\in\{\mathcal{S,T}\}
\end{equation}
Here, $\delta_{\mathcal{Z}\mathcal{Z}^\prime}$ is the Kronecker delta, ensuring task-specific components only contribute to their own task's variance. As a result, the joint distribution of $\mathbf{f}_{\mathcal{S}}=[f_\mathcal{S}(\mathbf{x}_{\mathcal{S},1}),\dots,f_\mathcal{S}(\mathbf{x}_{\mathcal{S},M})]^\intercal$ and $\mathbf{f}_{\mathcal{T}}=[f_\mathcal{T}(\mathbf{x}_{\mathcal{T},1}),\dots,f_\mathcal{T}(\mathbf{x}_{\mathcal{T},N})]^\intercal$ is multivariate Gaussian:
\begin{equation}
\begin{bmatrix} \mathbf{f}_{\mathcal{S}} \\ \mathbf{f}_{\mathcal{T}}\end{bmatrix}
    \sim
    \mathcal{N}\left(\begin{bmatrix} \mathbf{0}_M \\ \mathbf{0}_N \end{bmatrix},
     \mathbf{K}\circ \mathbf{C} \right)
\end{equation}

where $\circ$ is the Hadamard product (element-wise product). $\mathbf{K}$ and $\mathbf{C}$ are:
\begin{equation}
    \mathbf{K} =
    \begin{bmatrix}
    k(\mathbf{X}_\mathcal{S},\mathbf{X}_\mathcal{S}) & k(\mathbf{X}_\mathcal{S},\mathbf{X}_\mathcal{T}) \\
    k(\mathbf{X}_\mathcal{T},\mathbf{X}_\mathcal{S}) & k(\mathbf{X}_\mathcal{T},\mathbf{X}_\mathcal{T})
    \end{bmatrix}, \quad
    \mathbf{C} =
    \begin{bmatrix}
    C_\mathcal{S,S}\mathbf{1}_{M\times M} & C_\mathcal{S,T}\mathbf{1}_{M\times N} \\
    C_\mathcal{T,S}\mathbf{1}_{N\times M} & C_\mathcal{T,T}\mathbf{1}_{N\times N}
    \end{bmatrix}
\end{equation}
where $k(\mathbf{X}_\mathcal{Z},\mathbf{X}_\mathcal{Z^\prime})$ is the covariance matrix, and $\mathbf{1}_{R\times H}$ is an $R \times H$ matrix of ones.

\subsection{Learning Task Relationships via Hierarchical Regularization}

The task embeddings $\mathbf{e}_\mathcal{S}$ and $\mathbf{e}_\mathcal{T}$ are critical. Their locations in the embedding space determine how much structural information transfers between the SABR synthetic data and market observations. However, optimizing these embeddings can be challenging when directly maximizing the marginal likelihood, especially with imbalanced datasets.

When the target dataset $\mathcal{D}_\mathcal{T}$ is much sparser than the source dataset $\mathcal{D}_\mathcal{S}$, standard maximum marginal likelihood estimation (MLE) can lead to two undesirable outcomes:
\begin{enumerate}
    \item The embeddings might diverge too much, effectively isolating the tasks and preventing useful knowledge transfer from the denser source task.
    \item Alternatively, the optimization might be dominated by the dense source data, forcing the embeddings too close together and leading to inappropriate information transfer that ignores potentially significant differences between SABR predictions and market data.
\end{enumerate}
Simply fixing a predetermined correlation level would be too restrictive. We need an adaptive approach that provides guidance while letting the data inform the level of information sharing.

We addressed this using a hierarchical Bayesian formulation within a maximum a posteriori (MAP) estimation framework. Instead of treating task embeddings as independent parameters to be learned solely through the likelihood, we assumed they are drawn from a common prior distribution, governed by hyperparameters $\bm{\mu}_{\text{e}}$ and $\sigma^2_{\text{e}}$:
\begin{equation}
    \label{eq:embedding_prior_dist}
    P(\mathbf{e}_\mathcal{Z} | \bm{\mu}_{\text{e}}, \sigma^2_{\text{e}}) = \mathcal{N}(\mathbf{e}_\mathcal{Z} | \bm{\mu}_{\text{e}}, \sigma^2_{\text{e}} \mathbf{I}_{d^\prime}) \quad \text{for } \mathcal{Z} \in \{\mathcal{S}, \mathcal{T}\}.
\end{equation}
Here, $\bm{\mu}_{\text{e}} \in \mathbb{R}^{d^\prime}$ and $\sigma^2_{\text{e}}$ controls the squared distance between task embeddings and their mean. We learned both the embeddings $\mathbf{e}_\mathcal{S}, \mathbf{e}_\mathcal{T}$ and the hyperparameters $\bm{\mu}_{\text{e}}, \sigma^2_{\text{e}}$ from data.

This hierarchical prior was incorporated into the objective function. We aimed to maximize the posterior probability $P(\boldsymbol{\theta} | \mathbf{y}, \mathbf{X})$, where $\mathbf{y} = [\mathbf{y}_{\mathcal{S}}^\intercal, \mathbf{y}_{\mathcal{T}}^\intercal]^\intercal$ and $\boldsymbol{\theta}$ represented all model parameters (embeddings, kernel parameters, variances, hierarchical hyperparameters, etc.). By Bayes' theorem, $P(\boldsymbol{\theta} | \mathbf{y}, \mathbf{X}) \propto P(\mathbf{y} | \mathbf{X}, \boldsymbol{\theta}) P(\boldsymbol{\theta})$. Maximizing the posterior is equivalent to minimizing its negative logarithm:
\begin{equation}
    \label{eq:map_objective}
    \hat{\boldsymbol{\theta}}_{\text{MAP}} = \arg\min_{\boldsymbol{\theta}} \left[ -\log P(\mathbf{y} | \mathbf{X}, \boldsymbol{\theta}) - \log P(\boldsymbol{\theta}) \right]
\end{equation}
For the prior term $P(\boldsymbol{\theta})$, we focused on the hierarchical prior for the embeddings $P(\mathbf{e}_\mathcal{S}, \mathbf{e}_\mathcal{T} | \bm{\mu}_{\text{e}}, \sigma^2_{\text{e}})$. We assumed that other parameters in $\boldsymbol{\theta}$ have flat (uninformative) priors and do not affect the optimization minimum. Thus, the relevant negative log prior term is:
\begin{align}
    \label{eq:neg_log_prior_term}
    -\log P(\boldsymbol{\theta}) = &-\log P(\mathbf{e}_\mathcal{S}, \mathbf{e}_\mathcal{T} | \bm{\mu}_{\text{e}}, \sigma^2_{\text{e}})  \nonumber \\
    = &\sum_{\mathcal{Z} \in \{\mathcal{S}, \mathcal{T}\}} \left( \frac{\| \mathbf{e}_\mathcal{Z} - \bm\mu_{\text{e}} \|^2}{2\sigma^2_{\text{e}}} + \frac{d^\prime}{2} \log \sigma^2_{\text{e}} + \frac{d^\prime}{2} \log(2\pi) \right)
\end{align}

The negative log marginal likelihood $-\log P(\mathbf{y} | \mathbf{X}, \boldsymbol{\theta})$, derived from the joint Gaussian distribution of observations $\mathbf{y} \sim \mathcal{N}(\mathbf{0}, \mathbf{K}\circ \mathbf{C} + \mathbf{\Sigma})$ under the MTGP model, is given exactly by:
\begin{equation}
    \label{eq:nlml_term_full}
    -\log P(\mathbf{y} | \mathbf{X}, \boldsymbol{\theta}) = \frac{1}{2}\mathbf{y}^\intercal (\mathbf{K}\circ \mathbf{C}  + \mathbf{\Sigma})^{-1}\mathbf{y} + \frac{1}{2}\log|\mathbf{K}\circ \mathbf{C} + \mathbf{\Sigma}| + \frac{M+N}{2} \log(2\pi)
\end{equation}
where $\mathbf{K} \circ \mathbf{C}$ is the prior covariance matrix of the latent functions at training points ($M$ source, $N$ target), and $\bm\Sigma = \text{diag}(\sigma_\mathcal{S}^2 \mathbf{1}_M, \sigma_\mathcal{T}^2 \mathbf{1}_N)$ is the diagonal noise covariance matrix.

Substituting the negative log marginal likelihood (NLML, equation (\ref{eq:nlml_term_full})) and the relevant negative log prior (equation (\ref{eq:neg_log_prior_term})) into the MAP objective (equation (\ref{eq:map_objective})), we arrived at the final objective function $\mathcal{L}(\boldsymbol{\theta})$ to minimize:
\begin{align}
\mathcal{L}(\boldsymbol{\theta}) = & \underbrace{\frac{1}{2}\mathbf{y}^\intercal (\mathbf{K}\circ \mathbf{C}  + \mathbf{\Sigma})^{-1}\mathbf{y} + \frac{1}{2}\log|\mathbf{K}\circ \mathbf{C} + \mathbf{\Sigma}|}_{\text{Data Fidelity Term (NLML)}} \nonumber \\
& \underbrace{+ \sum_{\mathcal{Z} \in \{\mathcal{S}, \mathcal{T}\}} \left( \frac{\| \mathbf{e}_\mathcal{Z} - \bm\mu_{\text{e}} \|^2}{2\sigma^2_{\text{e}}} + \frac{d^\prime}{2} \log \sigma^2_{\text{e}} \right)}_{\text{Hierarchical Regularization Term (Neg. Log Prior)}}  \underbrace{+ \frac{M+N+2d^\prime}{2} \log(2\pi)}_{\text{Constant Term}} \label{eq:nll_revised_full}
\end{align}

We optimized equation (\ref{eq:nll_revised_full}) using L-BFGS to find:
\begin{itemize}
    \item Optimal task embeddings $\mathbf{e}_\mathcal{S}$ and $\mathbf{e}_\mathcal{T}$
    \item Hierarchical distribution parameters $\bm{\mu}_{\text{e}}$ and $\sigma^2_{\text{e}}$
    \item Other model hyperparameters (kernel parameters, noise variances, etc.)
\end{itemize}

The optimization balanced fitting the data well while keeping task embeddings plausible under their common prior. The regularization strength adapted based on the learned variance $\sigma^2_{\text{e}}$.

This approach offered advantages for volatility construction. It provided a data-driven way to determine information transfer without fixed assumptions. When SABR structure matches market patterns well, embeddings could move closer, increasing information flow. When differences exist, embeddings could maintain separation while still benefiting from shared characteristics encoded by the kernel. 

\subsection{Prediction}

Once we optimized all model parameters, we could make predictions for the target implied volatility surface at any new test point. The prediction followed standard GP principles, adapted for our multitask framework.

Given optimized parameters $\hat{\boldsymbol{\theta}}$ and the combined training dataset $\mathcal{D} = \mathcal{D}_\mathcal{S} \cup \mathcal{D}_\mathcal{T}$, we computed the predictive distribution for the target function $f_\mathcal{T}$ at a new test point $\mathbf{x}_\ast = [K_\ast, \tau_\ast]^\intercal$. This involved the joint Gaussian distribution of the target function value $f_\mathcal{T}(\mathbf{x}_\ast)$ and all training observations $\mathbf{y}$:

\begin{equation}
   \begin{bmatrix} \mathbf{y} \\ f_\mathcal{T}(\mathbf{x}_\ast)\end{bmatrix}
    \sim 
    \mathcal{N}\left(\begin{bmatrix} \mathbf{0}_{M+N} \\ 0 \end{bmatrix}, 
    \begin{bmatrix}
    \mathbf{K}\circ \mathbf{C} + \bm{\Sigma} & \mathbf{k}_\ast
    \\
    \mathbf{k}_\ast^{\intercal} & C_{\mathcal{T,T}} k(\mathbf{x}_{\ast},\mathbf{x}_{\ast})
    \end{bmatrix} \right)
\end{equation}

The cross-covariance vector $\mathbf{k}_\ast$ related the test point to all training points across both tasks, incorporating learned task relationships via $C_{\mathcal{T,S}}$ and $C_{\mathcal{T,T}}$:
\begin{equation}
\mathbf{k}_\ast=
\Big[
\underbrace{C_{\mathcal{T,S}}k(\mathbf{x}_\ast,\mathbf{x}_{\mathcal{S},1}),\dots,C_{\mathcal{T,S}}k(\mathbf{x}_\ast,\mathbf{x}_{\mathcal{S},M})}_{\text{Covariance with source (SABR) points}} ,\underbrace{C_{\mathcal{T,T}}k(\mathbf{x}_\ast,\mathbf{x}_{\mathcal{T},1}),\dots,C_{\mathcal{T,T}}k(\mathbf{x}_\ast,\mathbf{x}_{\mathcal{T},N})}_{\text{Covariance with target (market) points}}
\Big]^\intercal
\end{equation}

The factor $C_{\mathcal{T,S}}$ controlled the influence of synthetic SABR data, while $C_{\mathcal{T,T}}$ determined the weight of the market data. These factors reflected the learned task relationship.

Using standard GP conditioning formulas, the predictive distribution for the target task at $\mathbf{x}_\ast$ was Gaussian:
\begin{align}
f_\mathcal{T}(\mathbf{x}_\ast) | \mathcal{D} &\sim \mathcal{N}(\mu_\ast, \sigma_\ast^2)\\
\mu_\ast &= \mathbf{k}_\ast^\intercal(\mathbf{K}\circ \mathbf{C}  + \mathbf{\Sigma})^{-1}\mathbf{y} \label{eq:pred_mean}\\
\sigma_\ast^2 &= C_{\mathcal{T,T}} k(\mathbf{x}_{\ast}, \mathbf{x}_\ast) - \mathbf{k}_\ast^\intercal(\mathbf{K}\circ \mathbf{C}  + \mathbf{\Sigma})^{-1}\mathbf{k}_\ast \label{eq:pred_var}
\end{align}
The mean $\mu_\ast$ was our estimation of implied volatility at $\mathbf{x}^\ast$, and the variance $\sigma_\ast^2$ quantified prediction uncertainty.

\section{Experimental Design} \label{sec:experiments}

To evaluate SABR-MTGP against benchmark methods, we designed numerical experiments. Since the goal was to assess implied volatility surface construction accuracy, we required a reliable ground truth reflecting realistic market behavior. We chose the Heston stochastic volatility model \citep{heston1993closed} for this purpose, as it generates complex volatility dynamics similar to real markets.

Our experimental setup comprised the generation of ground-truth data, construction of sparse ``market'' observations, slice-wise SABR calibration, model training, and evaluation across market regimes. This section details the methodology, and Section~\ref{sec:results} presents the findings.

\subsection{Data Generation Framework}

We used two datasets: a sparse ``market'' dataset and a denser synthetic dataset obtained by SABR calibration on the market dataset. The market dataset serves as the target task, and the synthetic dataset serves as the source task for multitask learning.

\subsubsection{Heston Ground Truth Data}

We generated synthetic ``market''  data using the Heston model \citep{heston1993closed}, which describes asset price $S_t$ and variance $v_t$ dynamics via SDEs:
\begin{align}
dS_t &= (r-q)S_t dt + \sqrt{v_t} S_t dW_t^{(1)} \label{eq:heston_S}\\
dv_t &= \kappa(\theta - v_t)dt + \nu_{\text{vol}} \sqrt{v_t} dW_t^{(2)} \label{eq:heston_v}
\end{align}
Here, $r$ is the risk-free rate, $q$ the dividend yield, $\kappa$ the mean-reversion speed, $\theta$ the long-run variance, and $\nu_{\text{vol}}$ the volatility-of-volatility. Brownian motions $W_t^{(1)}$ and $W_t^{(2)}$ have correlation $\rho$.

We chose Heston for its ability to reproduce realistic smiles and skews. Its characteristic function based pricing enables high-precision computations for ground-truth generation. Moreover, its parameters have financial interpretations, enabling simulation of various market regimes. For our baseline scenario (the \textit{Base} case), we used parameters: $v_0 = 0.09$, $\theta = 0.09$, $\kappa = 1.0$, $\nu_{\text{vol}} = 0.8$, and $\rho = -0.8$. We set $S_0 = 100$, $r = 0.03$, and $q = 0.01$. 

\begin{figure}[htbp]
    \centering
    \includegraphics[width=0.5\textwidth]{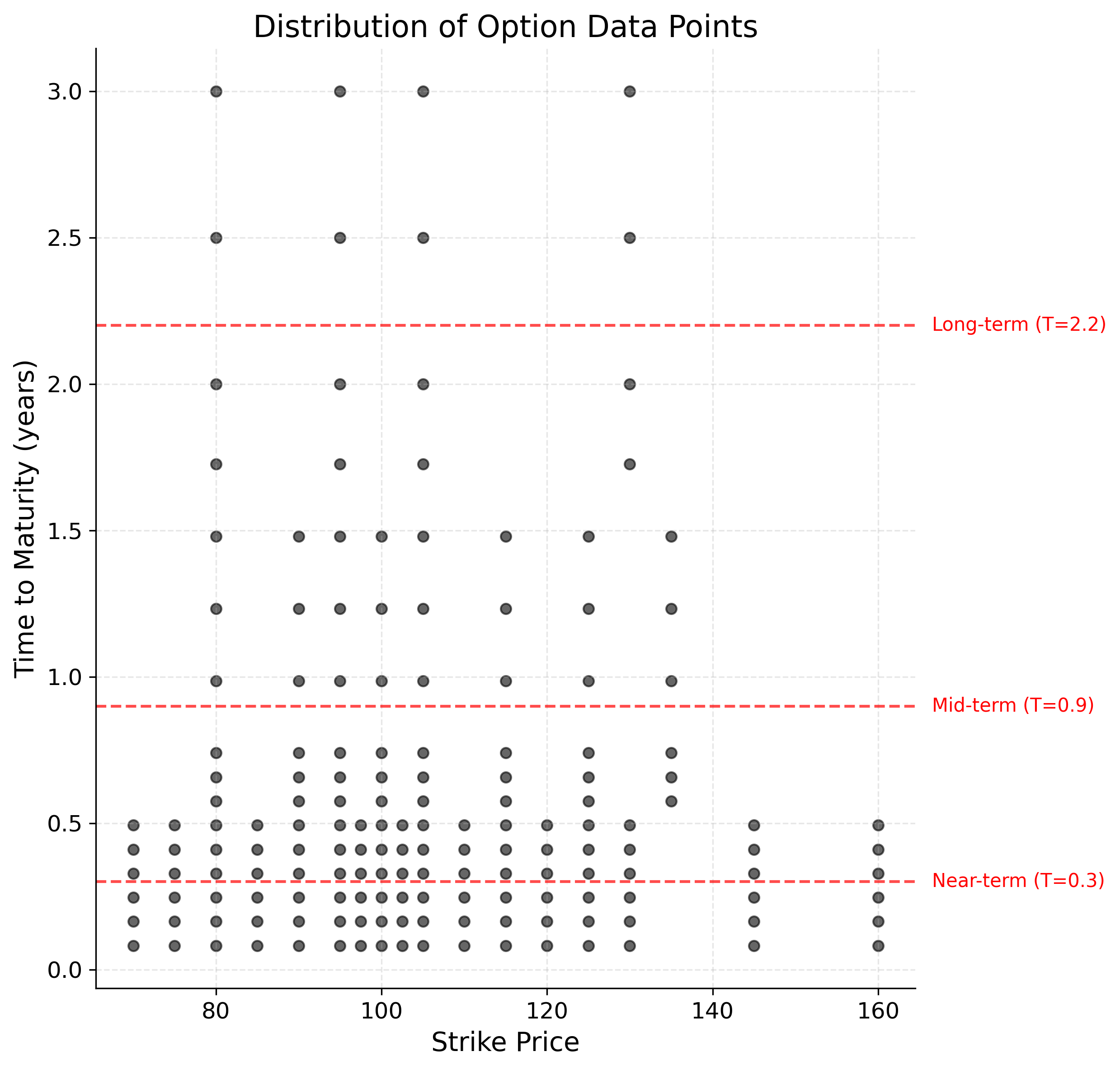}
    \caption{Distribution of generated Heston ``market'' data points (target task $\mathcal{D}_\mathcal{T}$) in our experiments, by strike price and time to maturity. Dashed red lines indicate evaluation maturities.}
    \label{fig:data_distribution}
\end{figure}

To simulate ``market'' option data, we generated a dataset $\mathcal{D}_\mathcal{T}$ comprising $N=166$ European call option contracts. We designed maturities to mirror market availability, which is dense at short horizons and sparser at longer horizons. This led to expirations about monthly up to 9 months (0.08 to 0.75 years). Then, we used sparser intervals close to quarterly steps up to 1.75 years. Additionally, we included annual expirations at 2.0, 2.5, and 3.0 years. For each maturity, we chose strike prices ($K$) to cover a specific moneyness range ($K/S_0$). This range depended on the maturity: $[0.7, 1.6]$ for short-term ($\tau \le 0.5$ years), and $[0.8, 1.4]$ for mid-term ($0.5 < \tau \le 1.5$ years) and long-term ($\tau > 1.5$ years). Inside these ranges, we changed the strike intervals in a planned way. Consequently, there were more strikes near the at-the-money (ATM) region (from $0.95 S_0$ to $1.05 S_0$) and fewer strikes further away. Specifically, the strike intervals around ATM, near ATM, and far from the money were (2.5, 5.0, 15.0) for short-term, (5.0, 10.0, 25.0) for mid-term, and (10.0, 25.0, 50.0) for long-term maturities. Figure~\ref{fig:data_distribution} shows the data point distribution for our experiments.

For each option, we computed the Heston price via the Carr-Madan fast Fourier transform method (FFT) \citep{carr1999option} using standard damping $\alpha=2.5$ and a high-resolution grid with $N=2^{21}$ and $c=12800$. We verified numerical stability by doubling $N$ and $c$, which changed call prices by less than $10^{-6}$. Here, $N$ denotes the number of frequency samples (FFT grid size) and $c$ is the integration-range truncation parameter in Carr-Madan FFT. The spacing $\eta=c/N$ and the log-strike span $\Lambda=\tfrac{2\pi}{N\eta}$ in the Carr-Madan formulation control the tradeoff between aliasing and accuracy. IVs are computed via a safeguarded Newton method (tolerance $10^{-5}$, maximum 200 iterations, $\sigma\in[10^{-4},5]$). This produced the ``market'' dataset $\mathcal{D}_\mathcal{T} = \{(\mathbf{X}_\mathcal{T}, \mathbf{y}_\mathcal{T})\}$, the target task data.

\subsubsection{SABR-generated Synthetic Data}
\begin{figure}[htbp]
    \centering
    \includegraphics[width=0.75\textwidth]{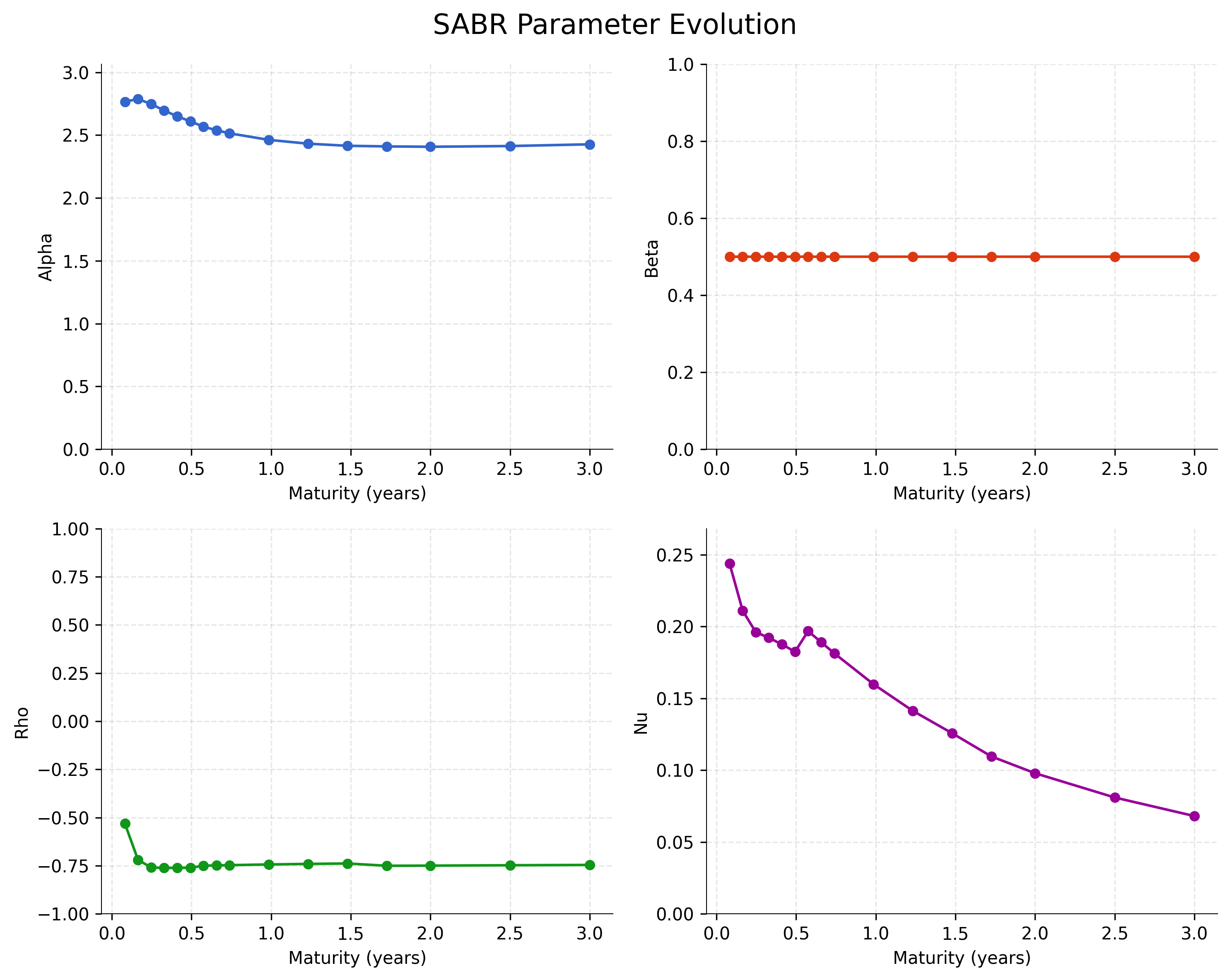}
    \caption{Evolution of calibrated SABR parameters ($\alpha, \beta, \rho, \nu$) across maturity slices for the \textit{Base} Heston scenario. Note $\beta$ was fixed at 0.5 during calibration.}
    \label{fig:sabr_params}
\end{figure}

The SABR-MTGP used a denser synthetic dataset derived by fitting the SABR model to the Heston data. We calibrated the SABR following the scheme in Section \ref{sec_SABR}. Figure~\ref{fig:sabr_params} shows the calibrated SABR parameters across maturities for the \textit{Base} Heston scenario.

Then, we defined a dense $35\times35$ grid $(\tilde{K}_j, \tilde{\tau}_j)$, covering strike range ($K \in [70, 160]$) and maturity range ($\tau \in [0.08, 3.0]$). At each grid point, we used the Hagan formula equation~(\ref{eq:sabr_detailed}) to compute synthetic implied volatility $y_{\mathcal{S},j} = \sigma_{\text{SABR}}(\tilde{K}_j, F_{\tilde{\tau}_j},\tilde{\tau}_j).$

Finally, we added small Gaussian noise $\epsilon_{\mathcal{S}} \sim \mathcal{N}(0, 0.01^2)$ to the synthetic data. This generated the synthetic source dataset $\mathcal{D}_\mathcal{S} = \{(\mathbf{X}_\mathcal{S}, \mathbf{y}_\mathcal{S})\}$.

\subsection{Comparative Models and Training}

We evaluated six methods: (i) \textbf{SABR}\footnote{Hagan asymptotics with fixed $\beta=0.5$; slice-wise calibration of $(\alpha,\rho,\nu)$.}, (ii) \textbf{Single-Task GP}\footnote{Matérn 5/2 kernel on $(K,\tau)$ with L-BFGS training}, (iii) the proposed \textbf{SABR-Informed Multitask GP (SABR-MTGP)}\footnote{ICM structure with a shared Matérn 5/2 input kernel and a task kernel based on learned one-dimensional embeddings; optimized by L-BFGS.}, (iv) arbitrage-free \textbf{SSVI} \citep{gatheral2014arbitrage}, (v) \textbf{Cubic Spline}, and (vi) \textbf{Neural Network with SABR pre-training}\footnote{Two-hidden-layer MLP (64 units each, ReLU) pre-trained on synthetic SABR data and fine-tuned on target data with Adam optimizer.}.

\subsection{Evaluation Framework and Robustness Analysis}

We evaluated model performance against the Heston ground truth on a predefined evaluation grid for a fair comparison. This grid, separate from training data, included points $(K_\ast, \tau_\ast)$ spanning a relevant moneyness range (e.g., $K_\ast/S_0 \in [0.8, 1.4]$) at specific test maturities $\tau_\ast$ (e.g., 0.3, 0.9, 2.2 years) representing near-, mid-, and long-term. For each point, we computed Heston prices via the Carr-Madan FFT and inverted to IVs by safeguarded Newton.

Each model (SABR, GP, SABR-MTGP, SSVI, Cubic Spline, and NN) predicted $\hat{\sigma}(K_\ast, \tau_\ast)$ on this grid. We measured performance using Root Mean Squared Error (RMSE) and Mean Absolute Error (MAE) against the ground truth $\sigma_{\text{Heston}}(K_\ast, \tau_\ast)$, calculated over valid predictions.

To assess robustness, we repeated experiments using ten Heston parameter configurations (Table~\ref{tab:heston_params}), including the \textit{Base} scenario. These settings simulated various market dynamics by altering parameters like mean reversion $\kappa$, volatility levels $\theta, \nu_{\text{vol}}$, correlation $\rho$, and initial term structure $v_0$ vs $\theta$. 

\begin{table}[htbp]
    \centering
    \caption{Heston Parameter Configurations for Robustness Analysis. Common parameters are $S_0=100, r=0.03, q=0.01$.}
    \label{tab:heston_params}
    \begin{tabular}{@{}lccccc@{}}
        \toprule
        Setting Name & $\kappa$ & $\theta$ & $\nu_{\text{vol}}$ & $\rho$ & $v_0$ \\
        \midrule
        Base                 & 1.0 & 0.09 & 0.8 & -0.8 & 0.09 \\
        Moderate Mean-Rev    & 2.0 & 0.09 & 0.8 & -0.8 & 0.09 \\
        Low Mean-Rev         & 0.5 & 0.09 & 0.8 & -0.8 & 0.09 \\
        High Vol-Regime      & 1.0 & 0.16 & 0.9 & -0.8 & 0.16 \\
        Low Vol-Regime       & 1.0 & 0.04 & 0.4 & -0.8 & 0.04 \\
        Moderate Correlation & 1.0 & 0.09 & 0.8 & -0.5 & 0.09 \\
        Strong Correlation   & 1.0 & 0.09 & 0.8 & -0.9 & 0.09 \\
        Term Structure Up    & 1.0 & 0.16 & 0.8 & -0.8 & 0.09 \\
        Term Structure Down  & 1.0 & 0.04 & 0.8 & -0.8 & 0.16 \\
        Mixed Regime         & 1.5 & 0.12 & 0.6 & -0.6 & 0.12 \\
        \bottomrule
    \end{tabular}
\end{table}

\section{Results and Discussion} \label{sec:results}

This section reports cross regime robustness and then uses Base scenario slices as diagnostics. We compare SABR, the single-task GP, SABR-MTGP, SSVI, cubic spline, and a neural network baseline.

\subsection{Performance and Robustness Across Market Conditions}

Tables~\ref{tab:robust_near} and \ref{tab:robust_long} summarize model accuracy (RMSE and MAE) across ten Heston settings at near-term and long-term horizons. Values are scaled by $10^{-3}$, and the best metric value per setting is shown in bold. At the short maturity, SABR-MTGP is best in eight settings and GP leads in two. At the long maturity, the leader alternates by regime, with SABR best in six settings, SABR-MTGP best in three, and GP best in one. SSVI and cubic spline are competitive in a few settings yet they are not consistently dominant. The neural network baseline is generally weaker.

\begin{table}[htbp]
    \centering
    
    \caption{Error summary at Near-term maturity ($\tau=0.3$). Values are scaled ($\times 10^{-3}$).}
    \label{tab:robust_near}
    \resizebox{\textwidth}{!}{%
    \begin{tabular}{@{}l*{6}{cc}@{}}
        \toprule
        Setting & \multicolumn{2}{c}{SABR} & \multicolumn{2}{c}{GP} & \multicolumn{2}{c}{SABR-MTGP} & \multicolumn{2}{c}{SSVI} & \multicolumn{2}{c}{Cubic Spline} & \multicolumn{2}{c}{NN} \\
        & RMSE & MAE & RMSE & MAE & RMSE & MAE & RMSE & MAE & RMSE & MAE & RMSE & MAE \\
        \midrule
        Base & 5.78 & 5.22 & 1.93 & 1.63 & \textbf{1.69} & \textbf{1.44} & 4.27 & 3.42 & 2.11 & 1.80 & 4.67 & 3.49 \\
        Moderate Mean-Rev & 4.50 & 4.04 & 1.55 & 1.29 & \textbf{1.42} & \textbf{1.25} & 4.09 & 3.17 & 1.79 & 1.54 & 6.31 & 4.36 \\
        Low Mean-Rev & 6.60 & 5.97 & 2.24 & \textbf{1.79} & \textbf{2.09} & 1.83 & 4.34 & 3.39 & 2.28 & 1.92 & 4.91 & 3.75 \\
        High Vol-Regime & 4.99 & 4.56 & 0.83 & 0.69 & \textbf{0.71} & \textbf{0.61} & 3.17 & 2.50 & 1.44 & 1.10 & 2.67 & 1.98 \\
        Low Vol-Regime & 6.74 & 6.36 & 4.15 & 2.48 & \textbf{3.94} & \textbf{2.35} & 4.04 & 3.15 & 5.04 & 3.17 & 10.89 & 7.37 \\
        Moderate Correlation & 4.10 & 3.72 & 1.38 & 1.08 & \textbf{1.23} & \textbf{1.05} & 1.90 & 1.71 & 1.93 & 1.45 & 3.21 & 2.66 \\
        Strong Correlation & 7.26 & 6.13 & 4.28 & 3.25 & \textbf{3.10} & \textbf{2.41} & 7.04 & 5.44 & 3.84 & 3.03 & 11.51 & 7.41 \\
        Term Structure Up & 4.88 & 4.37 & \textbf{1.34} & \textbf{1.14} & 1.59 & 1.41 & 4.42 & 3.34 & 1.84 & 1.57 & 5.02 & 3.60 \\
        Term Structure Down & 5.24 & 4.76 & 1.12 & 0.96 & \textbf{0.82} & \textbf{0.72} & 6.25 & 4.75 & 1.43 & 1.09 & 3.57 & 2.43 \\
        Mixed Regime & 2.16 & 2.00 & \textbf{0.38} & \textbf{0.29} & 0.89 & 0.65 & 1.12 & 0.83 & 0.75 & 0.56 & 2.03 & 1.52 \\
        \bottomrule
    \end{tabular}}
\end{table}

\begin{table}[htbp]
    \centering
    \small
    \caption{Error summary at Long-term maturity ($\tau=2.2$). Values are scaled ($\times 10^{-3}$).}
    \label{tab:robust_long}
    \resizebox{\textwidth}{!}{%
    \begin{tabular}{@{}l*{6}{cc}@{}}
        \toprule
        Setting & \multicolumn{2}{c}{SABR} & \multicolumn{2}{c}{GP} & \multicolumn{2}{c}{SABR-MTGP} & \multicolumn{2}{c}{SSVI} & \multicolumn{2}{c}{Cubic Spline} & \multicolumn{2}{c}{NN} \\
        & RMSE & MAE & RMSE & MAE & RMSE & MAE & RMSE & MAE & RMSE & MAE & RMSE & MAE \\
        \midrule
        Base & \textbf{0.42} & \textbf{0.27} & 0.82 & 0.69 & 0.61 & 0.59 & 0.45 & 0.33 & 1.08 & 0.66 & 4.60 & 4.10 \\
        Moderate Mean-Rev & \textbf{0.36} & \textbf{0.26} & 1.18 & 0.99 & 0.86 & 0.72 & 0.83 & 0.73 & 0.39 & 0.27 & 3.44 & 2.87 \\
        Low Mean-Rev & \textbf{0.55} & \textbf{0.45} & 1.60 & 1.17 & 1.81 & 1.56 & 1.60 & 1.32 & 3.15 & 1.96 & 5.24 & 4.48 \\
        High Vol-Regime & 0.33 & 0.24 & 0.24 & 0.19 & \textbf{0.17} & \textbf{0.14} & 0.90 & 0.76 & 0.43 & 0.31 & 4.05 & 3.53 \\
        Low Vol-Regime & 0.51 & \textbf{0.32} & 2.52 & 1.73 & \textbf{0.38} & 0.33 & 0.75 & 0.59 & 1.17 & 0.72 & 3.66 & 2.90 \\
        Moderate Correlation & \textbf{0.12} & \textbf{0.10} & 1.24 & 1.19 & 0.35 & 0.27 & 0.65 & 0.59 & 1.18 & 0.79 & 4.99 & 4.30 \\
        Strong Correlation & 0.74 & 0.41 & 1.36 & 0.67 & \textbf{0.15} & \textbf{0.12} & 0.83 & 0.74 & 0.58 & 0.34 & 4.01 & 3.54 \\
        Term Structure Up & 0.42 & 0.29 & \textbf{0.32} & \textbf{0.20} & 0.95 & 0.83 & 0.49 & 0.40 & 0.49 & 0.32 & 3.92 & 3.25 \\
        Term Structure Down & \textbf{0.57} & 0.54 & 0.97 & 0.84 & 0.68 & \textbf{0.47} & 2.34 & 1.87 & 1.69 & 1.22 & 3.66 & 3.24 \\
        Mixed Regime & \textbf{0.10} & \textbf{0.07} & 0.49 & 0.46 & 0.13 & 0.11 & 0.40 & 0.34 & 0.34 & 0.25 & 4.53 & 3.96 \\
        \bottomrule
    \end{tabular}}
\end{table}

\begin{figure}
    \centering
    \includegraphics[width=0.85\linewidth]{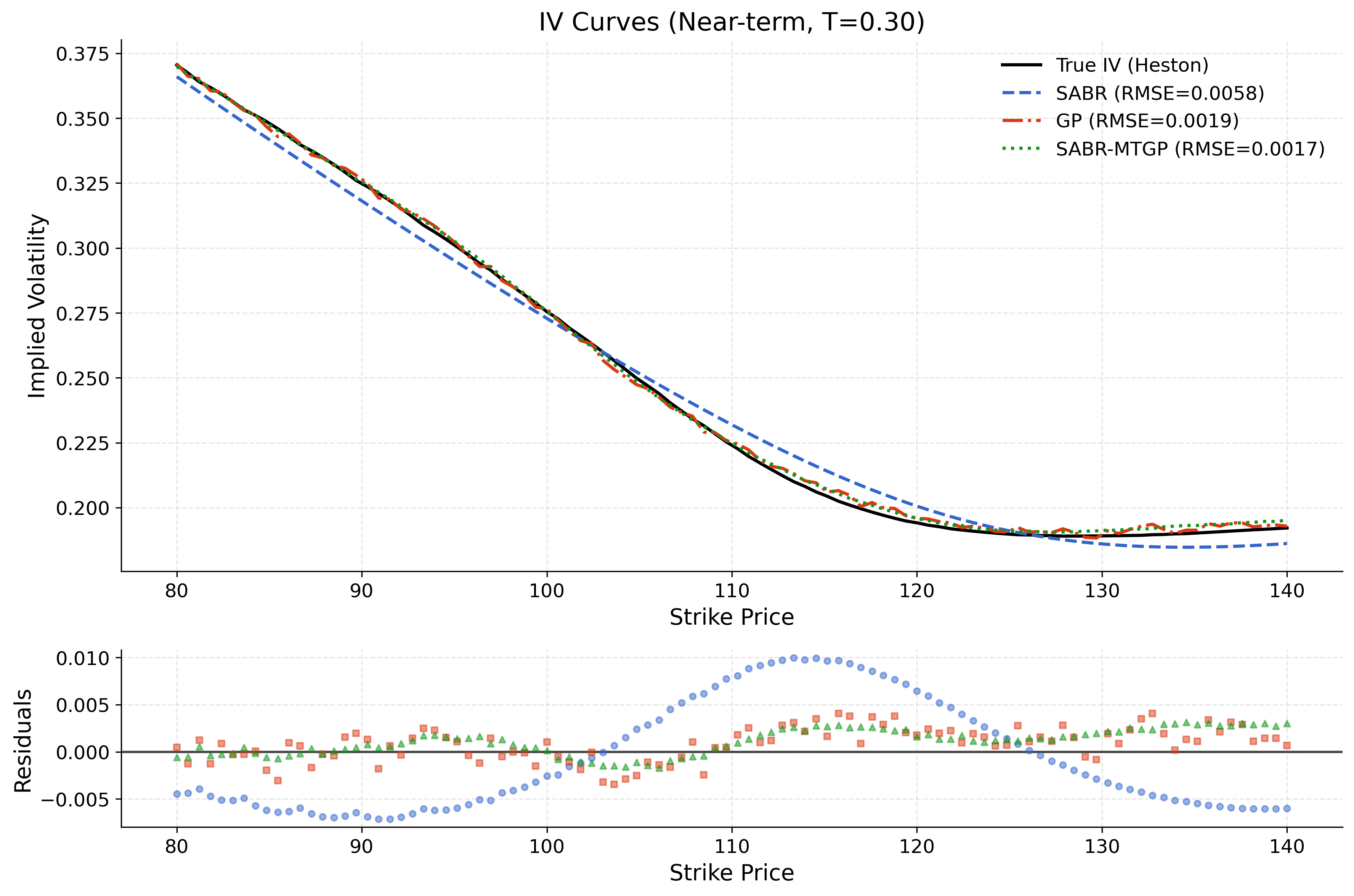}
    \caption{Implied volatility curve comparison and residuals under the \textit{Base} Heston scenario for near-term}
    \label{fig:iv_near}
\end{figure}

\begin{figure}
    \centering
    \includegraphics[width=0.85\linewidth]{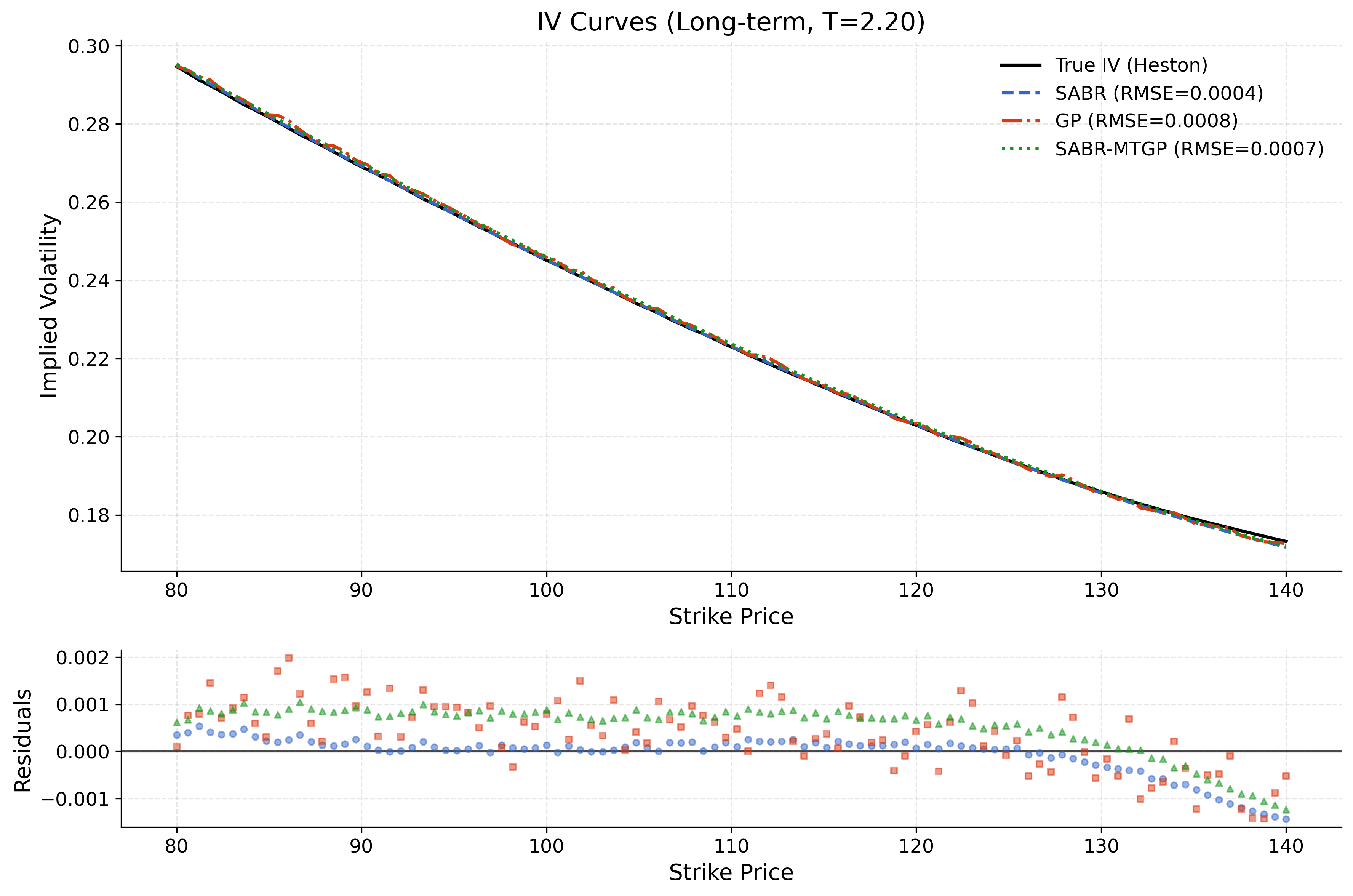}
    \caption{Implied volatility curve comparison and residuals under the \textit{Base} Heston scenario for long-term}
    \label{fig:iv_long}
\end{figure}

Figures \ref{fig:iv_near} and \ref{fig:iv_long} provide visual evidence from the Base scenario. In the near-term slice, SABR shows a systematic error pattern relative to the Heston truth that is consistent with the limitations of a rigid parametric form. The single-task GP follows the curve yet exhibits small local oscillations in the wings where observations are sparse. SABR-MTGP tracks the truth closely and its residuals are centered and smooth.


In the long-term slice, the three methods are closer. SABR is often the best while SABR-MTGP remains very close and smooth. The single-task GP shows mild undulations yet the deviations are small.

These patterns align with the design of SABR-MTGP. Short maturities contain more observations and the SABR-based source task acts as a stabilizer that suppresses small oscillations of a pure GP while preserving flexibility. At long maturities data are sparse and a simple parametric structure can be competitive. SABR-MTGP uses learned task embeddings with hierarchical regularization to adapt the level of transfer. When SABR and the target regime align, the learned correlation increases (cf. Section \ref{sec_task}).

\subsection{No-Arbitrage Analysis}\label{sec_noarb}
\begin{figure}[htbp]
    \centering
    \begin{subfigure}[b]{\textwidth}
        \centering
        \includegraphics[width=\textwidth]{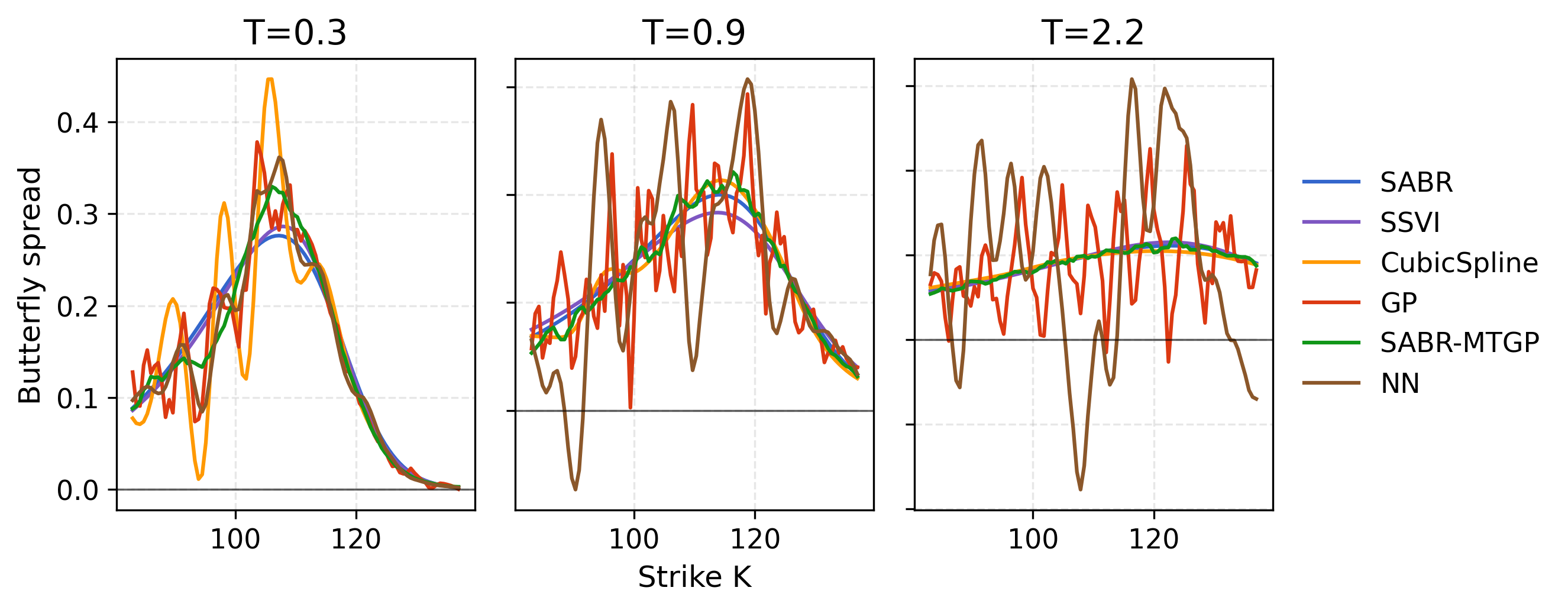}
        \caption{Butterfly spreads across strikes for three maturities (Base).}
        \label{fig:butterfly_base}
    \end{subfigure}
    \hfill
    \\
    \begin{subfigure}[b]{\textwidth}
        \centering
        \includegraphics[width=\textwidth]{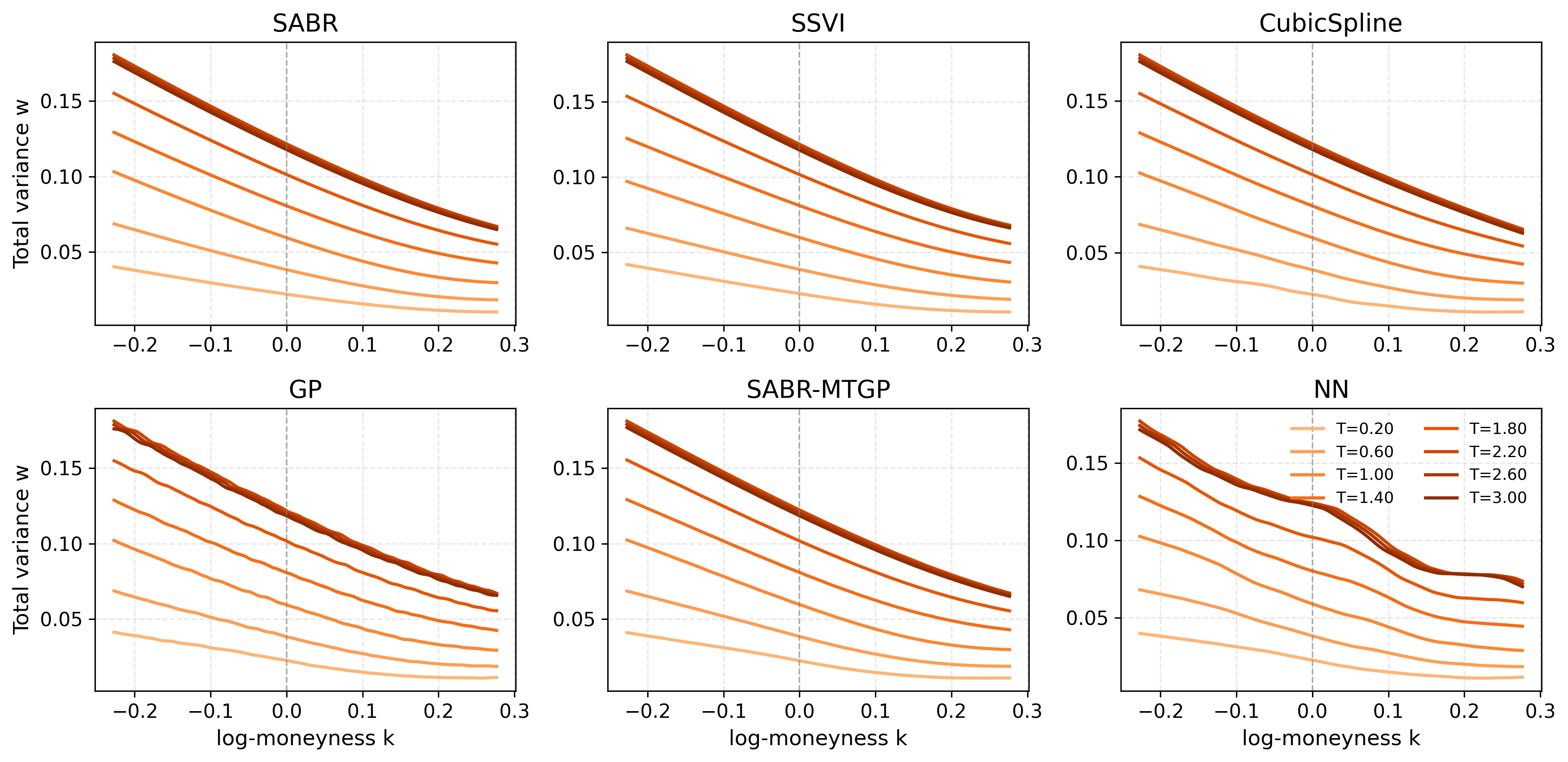}
        \caption{Total variance $w(k,\tau)$ across methods (six panels, Base).}
        \label{fig:tv_mtgp_only}
    \end{subfigure}
    \caption{No-arbitrage test. Panel (a) visualizes butterfly spreads; most methods remain above the zero line within a small tolerance band, while NN and GP occasionally dip negative at mid/long maturities. Panel (b) shows six panels of total variance $w(k,\tau)$ curves for SABR, SSVI, Cubic Spline, GP, SABR-MTGP, and NN; for most methods the level of $w$ increases with maturity at fixed $k$ and the $k$-profiles are smooth, with minor departures mainly for NN and GP.}
    \label{fig:noarb_viz}
\end{figure}

We evaluate no-arbitrage by computing the butterfly spread and total variance. For each maturity $\tau$ and ordered strikes $\{K_i\}_{i=1}^{100}\subset[80,140]$, we define the butterfly spread
\begin{equation}
    B_i(\tau) \,=\, C(K_{i+1},\tau) \, - \, 2\,C(K_i,\tau) \, + \, C(K_{i-1},\tau),
\end{equation}
which must satisfy $B_i(\tau)\ge 0$ (convexity in $K$). We report the butterfly violation rate as $\#\{i: B_i(\tau) < -\varepsilon\}/\#\{i\}$ with tolerance $\varepsilon = 2\times 10^{-5}$.  We define the total variance $w(k,\tau)=\sigma^2(k,\tau)\,\tau$ with forward $F_\tau$ and log-moneyness $k=\ln(K/F_\tau)$. The total variance should be non-decreasing in $\tau$ at fixed $k$ if the IVS is calendar arbitrage-free. Tables~\ref{tab:noarb_short}--\ref{tab:noarb_long} report violation rates, and Figure~\ref{fig:noarb_viz} visualizes butterfly spreads and total variance. Across Tables~\ref{tab:noarb_short}--\ref{tab:noarb_long}, SABR, SSVI, cubic spline, GP, and SABR-MTGP rarely exhibit butterfly violations on the evaluation grids; the few exceptions are small (\(\le 3.33\%\))  and arise mainly for GP at long-term maturity and for cubic spline at near-term maturity. The NN baseline shows higher violation rates in several regimes, especially at long-term maturities. Figure~\ref{fig:noarb_viz} corroborates these findings: butterfly spreads of the proposed method are nonnegative, and total variance panels exhibit the expected increase in $w$ with maturity at fixed $k$ consistent with the absence of calendar arbitrage.

\begin{table}[htbp]
    \centering
    \caption{Near-term butterfly violation rate (\(\tau=0.3\)). $\cross$ denotes no violation.}
    \label{tab:noarb_short}
     \resizebox{\textwidth}{!}{%
    {\small
    \begin{tabular}{@{}lcccccc@{}}
        \toprule
        Setting & SABR & SSVI & Cubic Spline & GP & SABR-MTGP & NN \\
        \midrule
        Base & $\cross$ & $\cross$ & $\cross$ & $\cross$ & $\cross$ & $\cross$ \\
        Moderate Mean-Rev & $\cross$ & $\cross$ & $\cross$ & $\cross$ & $\cross$ & $\cross$ \\
        Low Mean-Rev & $\cross$ & $\cross$ & 1.11\% & $\cross$ & $\cross$ & $\cross$ \\
        High Vol-Regime & $\cross$ & $\cross$ & $\cross$ & $\cross$ & $\cross$ & $\cross$ \\
        Low Vol-Regime & $\cross$ & $\cross$ & $\cross$ & $\cross$ & $\cross$ & 4.44\% \\
        Moderate Correlation & $\cross$ & $\cross$ & $\cross$ & $\cross$ & $\cross$ & $\cross$ \\
        Strong Correlation & $\cross$ & $\cross$ & 2.22\% & $\cross$ & $\cross$ & 1.11\% \\
        Term Structure Up & $\cross$ & $\cross$ & $\cross$ & $\cross$ & $\cross$ & 13.33\% \\
        Term Structure Down & $\cross$ & $\cross$ & $\cross$ & $\cross$ & $\cross$ & $\cross$ \\
        Mixed Regime & $\cross$ & $\cross$ & $\cross$ & $\cross$ & $\cross$ & $\cross$ \\
        \bottomrule
    \end{tabular}}
    }
\end{table}

\begin{table}[htbp]
    \centering
    \caption{Long-term butterfly violation rate (\(\tau=2.2\)). $\cross$ denotes no violation.}
    \label{tab:noarb_long}
    \resizebox{\textwidth}{!}{%
    {\small
    \begin{tabular}{@{}lcccccc@{}}
        \toprule
        Setting & SABR & SSVI & Cubic Spline & GP & SABR-MTGP & NN \\
        \midrule
        Base & $\cross$ & $\cross$ & $\cross$ & 3.33\% & $\cross$ & 25.56\% \\
        Moderate Mean-Rev & $\cross$ & $\cross$ & $\cross$ & $\cross$ & $\cross$ & 26.67\% \\
        Low Mean-Rev & $\cross$ & $\cross$ & $\cross$ & 1.11\% & $\cross$ & 22.22\% \\
        High Vol-Regime & $\cross$ & $\cross$ & $\cross$ & $\cross$ & $\cross$ & 37.78\% \\
        Low Vol-Regime & $\cross$ & $\cross$ & $\cross$ & $\cross$ & $\cross$ & 14.44\% \\
        Moderate Correlation & $\cross$ & $\cross$ & $\cross$ & $\cross$ & $\cross$ & 33.33\% \\
        Strong Correlation & $\cross$ & $\cross$ & $\cross$ & $\cross$ & $\cross$ & 24.44\% \\
        Term Structure Up & $\cross$ & $\cross$ & $\cross$ & $\cross$ & $\cross$ & 33.33\% \\
        Term Structure Down & $\cross$ & $\cross$ & $\cross$ & 1.11\% & $\cross$ & 21.11\% \\
        Mixed Regime & $\cross$ & $\cross$ & $\cross$ & $\cross$ & $\cross$ & 35.56\% \\
        \bottomrule
    \end{tabular}}
    } 
\end{table}

\subsection{Learned Task Relationship}\label{sec_task}
\begin{figure}[htbp]
    \centering
\begin{subfigure}[b]{\textwidth}
        \centering
        \includegraphics[width=\textwidth]{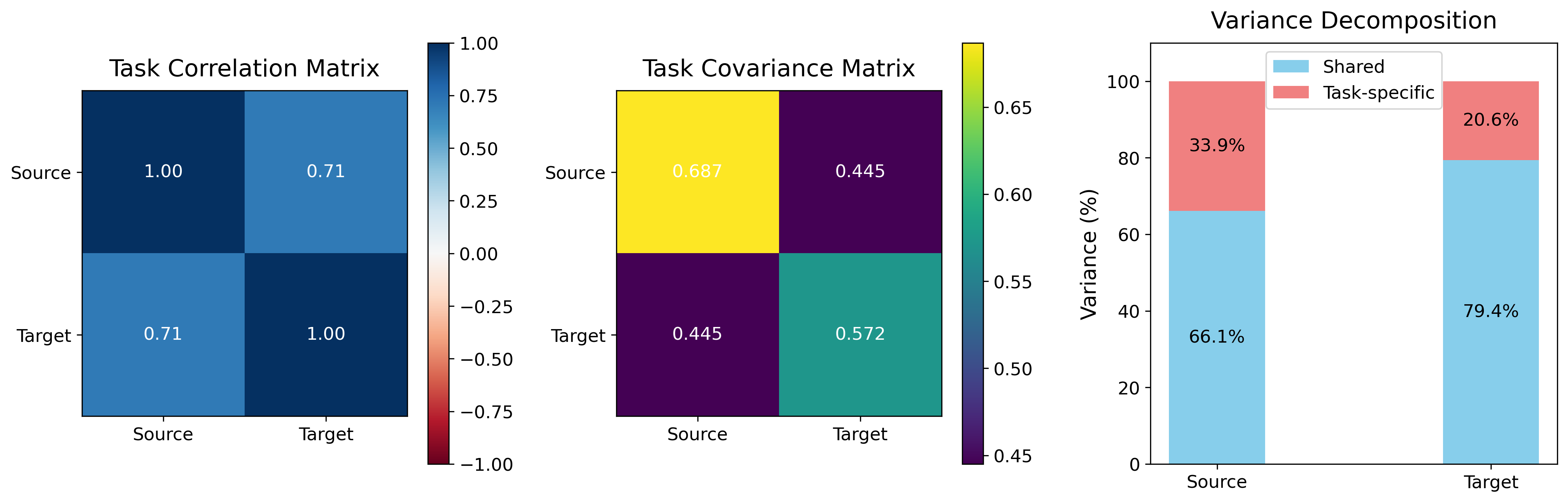}
        \caption{\textit{Term Structure Down} Setting.}
        \label{fig:task_analysis_low}
    \end{subfigure}
    \vfill
    \begin{subfigure}[b]{\textwidth}
        \centering
        \includegraphics[width=\textwidth]{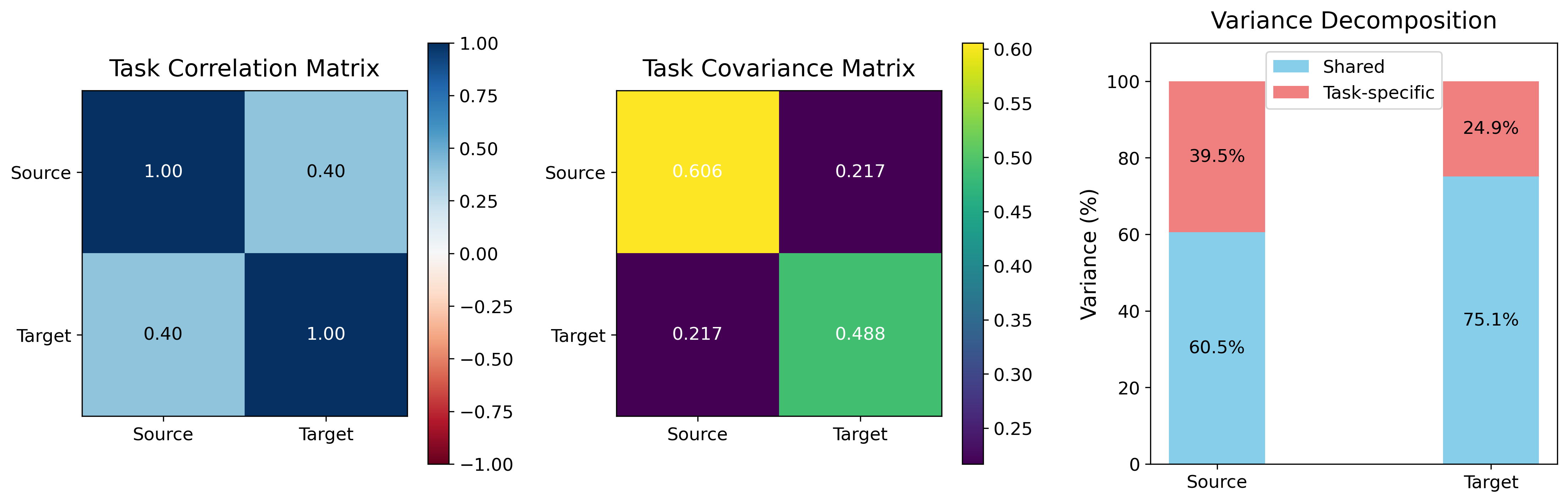}
        \caption{\textit{Mixed Regime} setting.}
        \label{fig:task_analysis_mod}
    \end{subfigure}

    \caption{Learned task relationship for two representative Heston settings: (a) \textit{Term Structure Down} and (b) \textit{Mixed Regime}. Each panel shows the task correlation, covariance, and variance decomposition. }
    \label{fig:task_analysis_examples}
\end{figure}

The MTGP framework learns the cross-task relationship between the SABR source and the Heston target. Figure~\ref{fig:task_analysis_examples} illustrates the learned task correlation, the task covariance, and the variance decomposition (task-specific proportion $\kappa^2_\mathcal{Z}/(\sigma^2_h+\kappa^2_\mathcal{Z})$ versus shared proportion $\sigma^2_h/(\sigma^2_h+\kappa^2_\mathcal{Z})$) of two representative settings.

In the \textit{Term Structure Down} case, the cross-task correlation is higher (\(\approx 0.71\)), indicating stronger transfer when the regime aligns with SABR's structure. In the \textit{Mixed Regime}, the lower correlation (\(\approx 0.40\)) implies greater reliance on task-specific variation. This adaptive behavior limits negative transfer while preserving the benefits of information transfer from SABR.

\subsection{Sensitivity Analyses}\label{sec:sensitivity}

\subsubsection{Noise in the source task}
\begin{figure}[t]
    \centering
    \includegraphics[width=\linewidth]{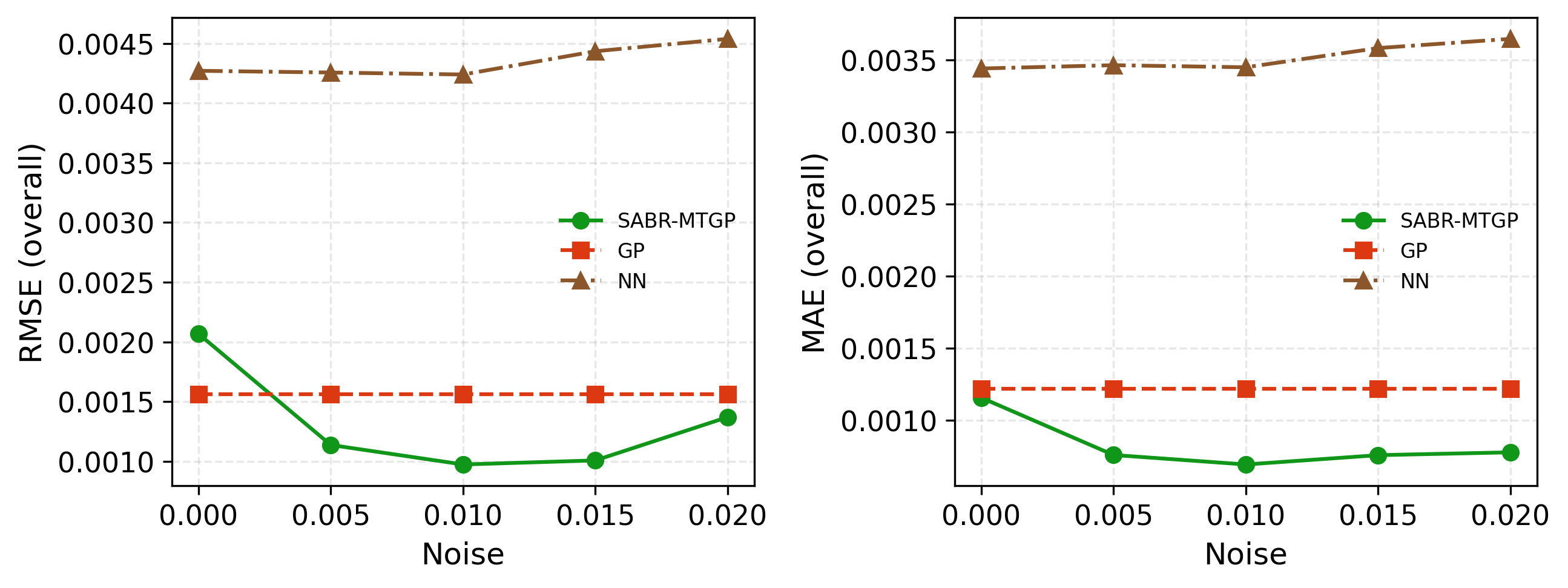}
    \caption{Sensitivity of overall accuracy to Gaussian noise injected into the SABR source labels. Left: RMSE vs.\ noise level $\sigma_\text{syn}$. Right: MAE vs.\ $\sigma_\text{syn}$. SABR-MTGP benefits from modest noise (minimum near $\sigma_\text{syn}\!\approx\!0.01$), and NN degrades as $\sigma_\text{syn}$ increases.}
    \label{fig:noise_sensitivity}
\end{figure}

We study how noise $\epsilon_\mathcal{S}\sim\mathcal{N}(0,\sigma_\text{syn}^2)$ in the SABR-generated source task affects model accuracy. We inject zero-mean Gaussian noise with standard deviation $\sigma_\text{syn} \in \{0,\,0.005,\,0.01,\,0.015,\,0.02\}$ into the synthetic source implied volatilities, and evaluate overall accuracy on the same out-of-sample grid used elsewhere in this section. Figure~\ref{fig:noise_sensitivity} reports the Root Mean Squared Error (RMSE) and Mean Absolute Error (MAE) aggregated over the evaluation grid for three methods: SABR-MTGP, single-task GP, and a neural network pretrained on the source and finetuned on the target (NN).

Empirically, SABR-MTGP exhibits a clear U-shaped response: small amounts of noise reduce error and the best performance is attained around $0.01$, after which errors slowly increase. This behavior is consistent with the role of mild perturbations as a regularizer that reduces over-reliance on the synthetic dataset, while excessive noise reduces useful structural information. The NN baseline degrades monotonically with increasing $\sigma_\text{syn}$, reflecting its higher sensitivity to the quality of the pretraining dataset. Practically, these results suggest that modest source-task noise (here, $\sigma_\text{syn}$ in the range $[0.005,\,0.015]$) can improve the robustness and generalization of SABR-MTGP.

\subsubsection{Backbone parameter $\beta$}\label{sec:beta_robustness}
\begin{figure}[t]
    \centering
    \includegraphics[width=\linewidth]{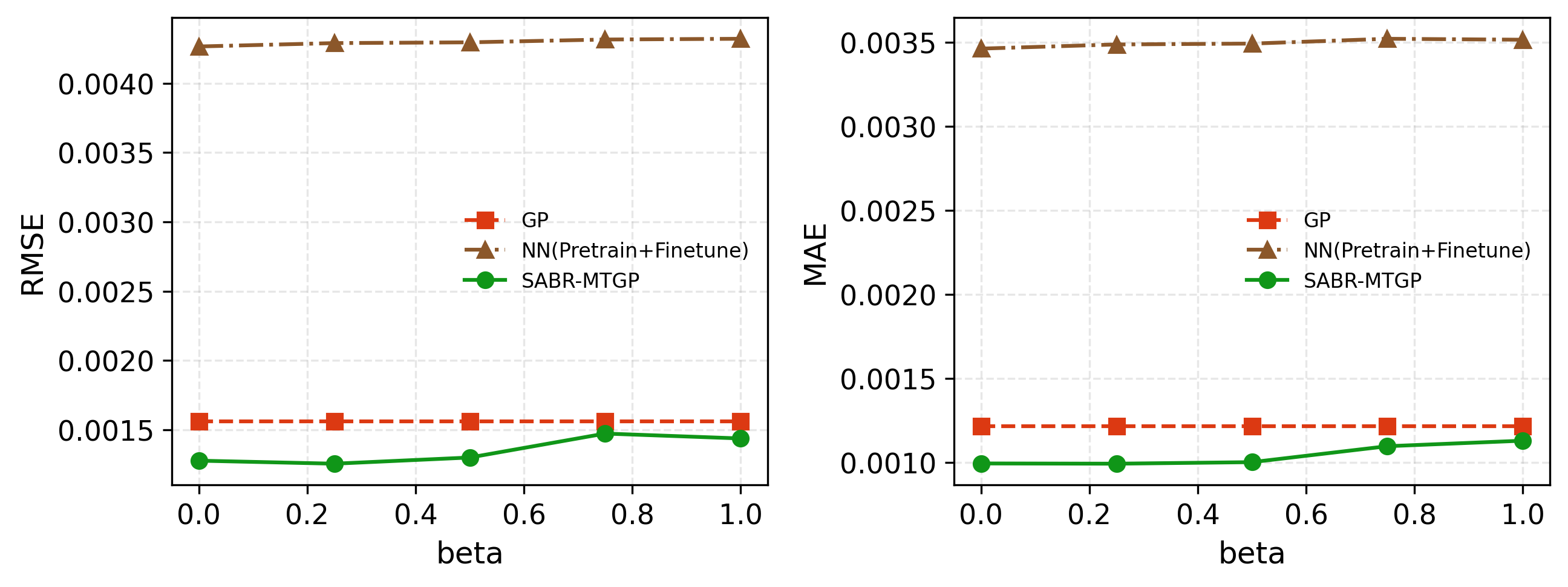}
    \caption{Effect of the SABR backbone parameter $\beta$ on overall accuracy in the Base setting. Left: RMSE vs.\ $\beta$. Right: MAE vs.\ $\beta$. }
    \label{fig:beta_robustness}
\end{figure}

In the previous sections, we treat the SABR backbone $\beta$ as a hyperparameter for generating synthetic data. In this section, we verify that our conclusions do not hinge on its value. We repeat the Base experiment for $\beta\in\{0,\,0.25,\,0.5,\,0.75,\,1.0\}$, and compute out-of-sample errors on the same evaluation grid.  Figure~\ref{fig:beta_robustness} shows that SABR-MTGP is insensitive to $\beta$ within a broad range, and outperforms GP in all cases. The neural network baseline is less accurate. 

\subsection{Summary}

Experiments on Heston data show that SABR-MTGP effectively constructs the IVS. By combining SABR's structural information with market observations in a multitask framework, it achieves better accuracy and stability than either GP or SABR alone, adaptively leveraging SABR insights according to their relevance to the target data. 

\section{Application to SPX Market Data} \label{sec:spx_application}

To assess the practical applicability of SABR-MTGP, we evaluate it on SPX option data from the \textit{OptionMetrics} database dated August 1, 2023. This case study complements the controlled experiments using the Heston model by assessing the method's behavior under actual market conditions.

\subsection{Data}

The dataset includes European call options on the SPX index. We preprocess the data by filtering for valid quotes, computing time-to-maturity, and computing implied volatilities from mid-prices. We use the zero-coupon yield curve for discounting and estimate the dividend yield via put-call parity. This gives us several hundred valid market implied volatility observations across different strikes and maturities up to about one year. Key statistics summarizing this dataset are in Table~\ref{tab:spx_data_summary}. This processed market dataset serves as the target task ($\mathcal{D}_\mathcal{T}$) for all methods.

\begin{table}[htbp]\small 
    \centering
    \caption{Descriptive Statistics for the Preprocessed SPX Call Option Dataset (August 1, 2023)}
    \label{tab:spx_data_summary}
    \begin{threeparttable}
    {\small 
        \begin{tabular}{@{}lc@{}}
            \toprule
            Statistic                 & Value \\
            \midrule
            Total Options             & 524 \\
            Unique Maturities         & 12 \\
            Maturity Range (Years)    & [0.0465, 0.9665] \\
            Strike Price Range        & [3300.00, 6600.00] \\
            Implied Volatility Range  & [0.0986, 0.3322] \\
            \midrule
            \multicolumn{2}{c}{Moneyness Distribution (Based on $K/S_0$, $S_0 = 4576.73$)} \\ 
            \midrule
            In-the-Money ($K/S_0 < 0.95$)   & 68 (13.0\%) \\
            At-the-Money ($0.95 \le K/S_0 \le 1.05$) & 289 (55.2\%) \\
            Out-of-the-Money ($K/S_0 > 1.05$)  & 167 (31.9\%) \\
            \bottomrule
        \end{tabular}
    } 
        \begin{tablenotes}
            \item Note: Statistics derived from the preprocessed SPX dataset used in this study. The moneyness definition uses $S_0 = 4576.73$.
        \end{tablenotes}
    \end{threeparttable}
\end{table}

\subsection{Results}

\begin{figure}[htbp]
    \centering
    \begin{subfigure}[b]{\textwidth}
        \centering
        \includegraphics[width=0.9\textwidth]{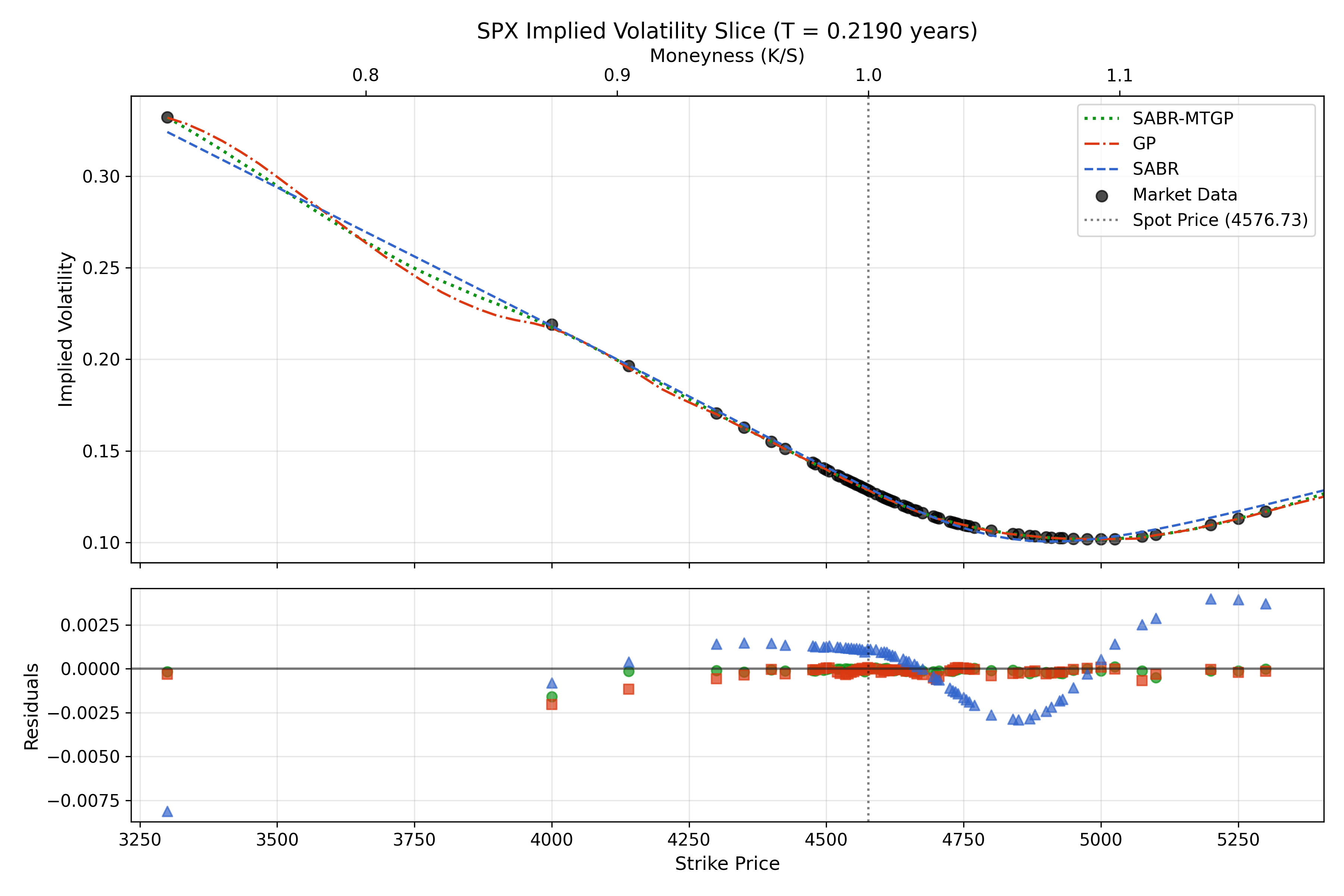} 
        \caption{Near-term maturity ($\tau = 0.2190$ years).}
        \label{fig:spx_slice_short_sub}
    \end{subfigure}
    \vfill 
    \begin{subfigure}[b]{\textwidth}
        \centering
        \includegraphics[width=0.9\textwidth]{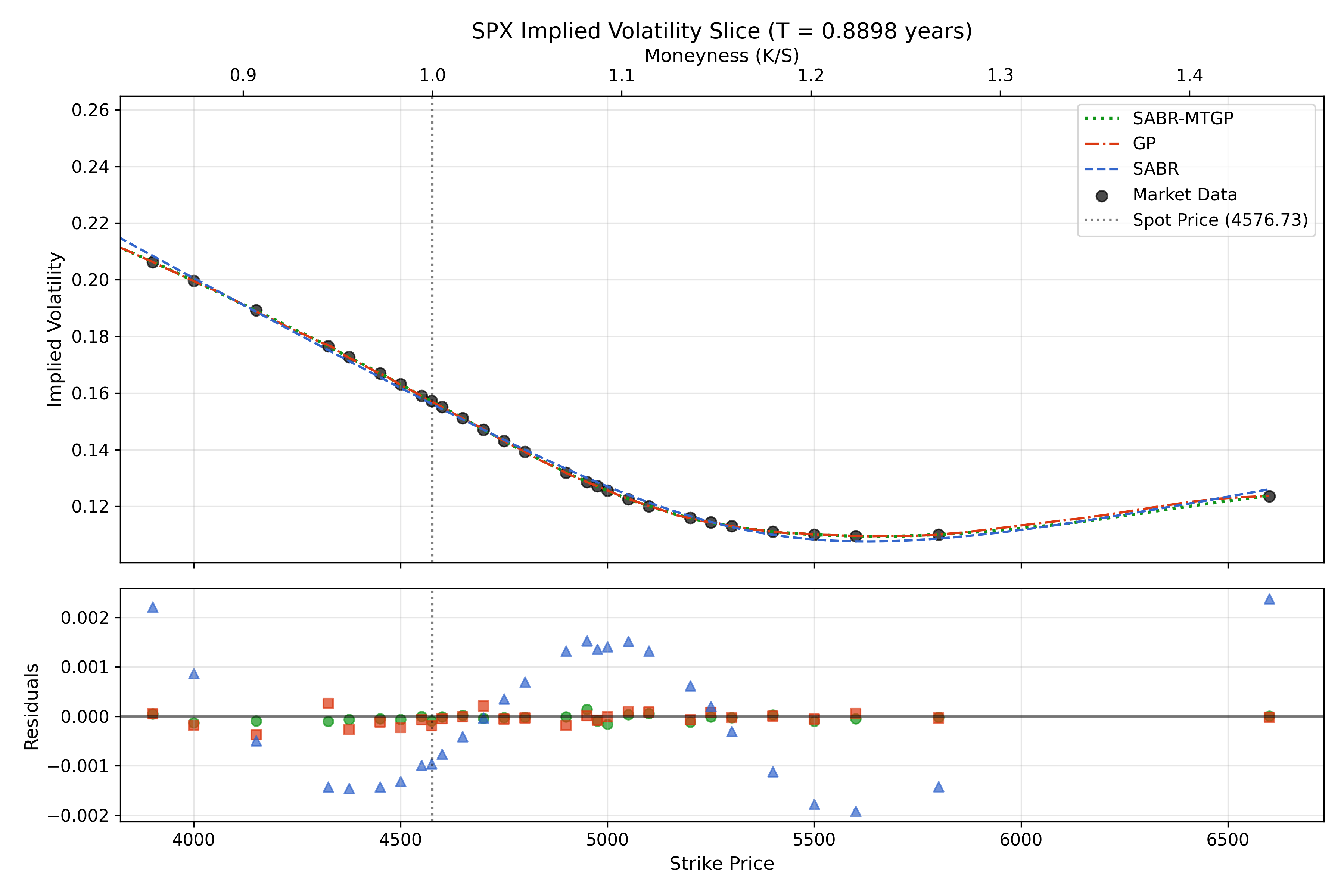} 
        \caption{Mid-term maturity ($\tau = 0.8898$ years).}
        \label{fig:spx_slice_mid_sub}
    \end{subfigure}
    \caption{Implied volatility slices for SPX options on August 1, 2023. Comparison of SABR-MTGP, GP, and SABR against market data points. Residuals (Model - Market) are shown below each slice.}
    \label{fig:spx_slices}
\end{figure}

Figure~\ref{fig:spx_slices} shows typical implied volatility slices for near-term ($\tau=0.2190$ years) and mid-term ($\tau=0.8898$ years) maturities.

In both slices (Figure~\ref{fig:spx_slices}), all three methods generally capture the observed smile/skew pattern of the market. However, the residual plots reveal notable differences in fit quality. For the near-term slice (Figure~\ref{fig:spx_slice_short_sub}), SABR has systematic errors shown by the consistent pattern in its residuals. Both the GP and SABR-MTGP fit the observed market points better, especially around the at-the-money region. However, there is a notable difference in the deep in-the-money region (strikes 3250-4000). Although downward-sloping, the standard GP curve exhibits local irregularities that render it less financially plausible than the SABR-implied curve. The SABR-MTGP curve maintains a smoother, decreasing volatility in this data-sparse region. This produces a more financially realistic interpolation, likely because of the structural guidance from the SABR source task. For the mid-term slice (Figure~\ref{fig:spx_slice_mid_sub}), where the smile is flatter and the market data is denser, all models achieve a very good fit. Notably, the SABR model fit improves significantly compared to the near-term slice. The residuals for both GP and SABR-MTGP are particularly small (generally within +/- 0.001), demonstrating excellent agreement with market observations.

\begin{figure}[]
    \centering
    \begin{subfigure}[b]{0.32\textwidth}
        \centering
        \includegraphics[width=\textwidth]{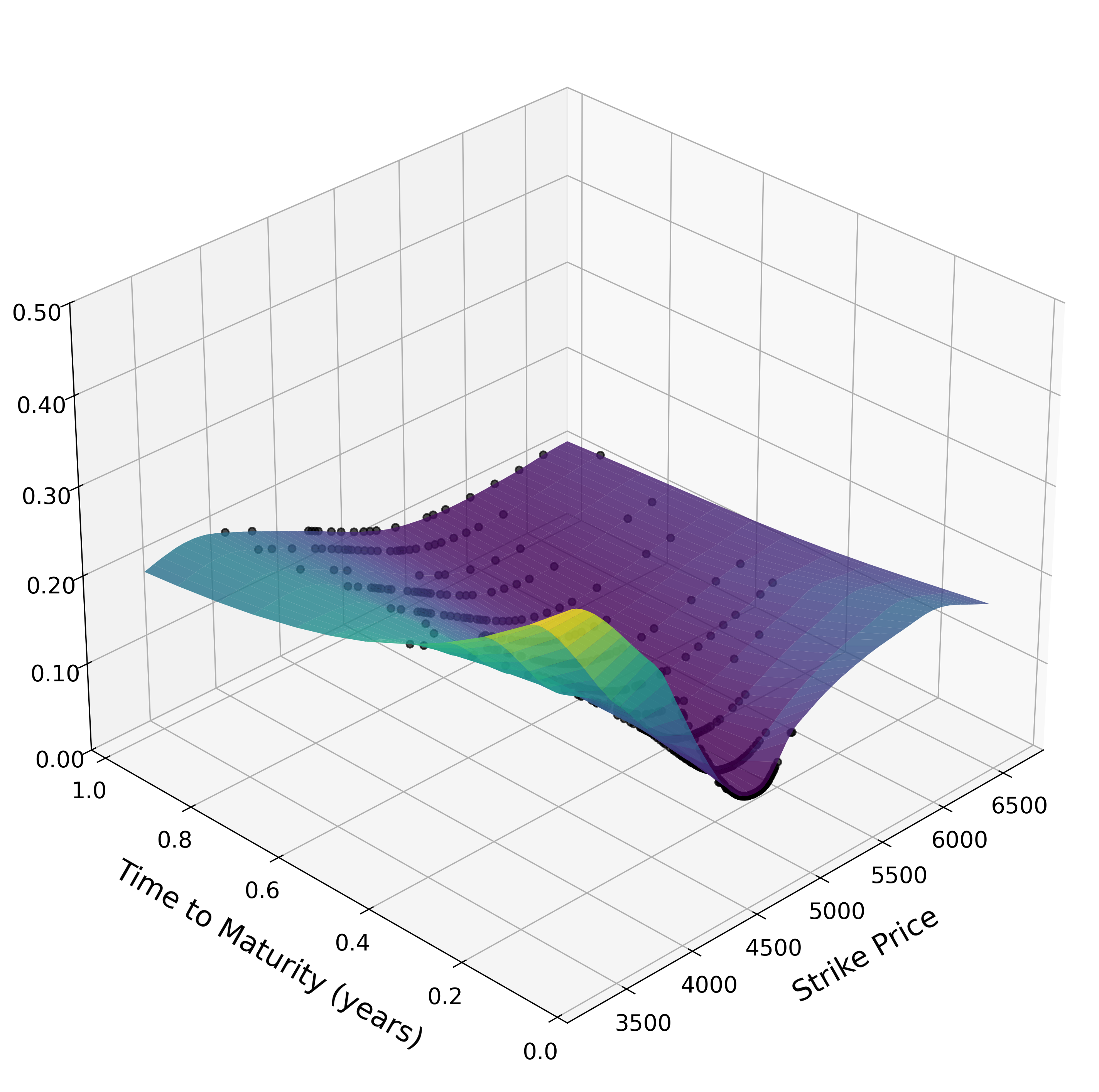}
        \caption{GP}
        \label{fig:spx_surface_gp_sub}
    \end{subfigure}
    \hfill
    \begin{subfigure}[b]{0.32\textwidth}
        \centering
        \includegraphics[width=\textwidth]{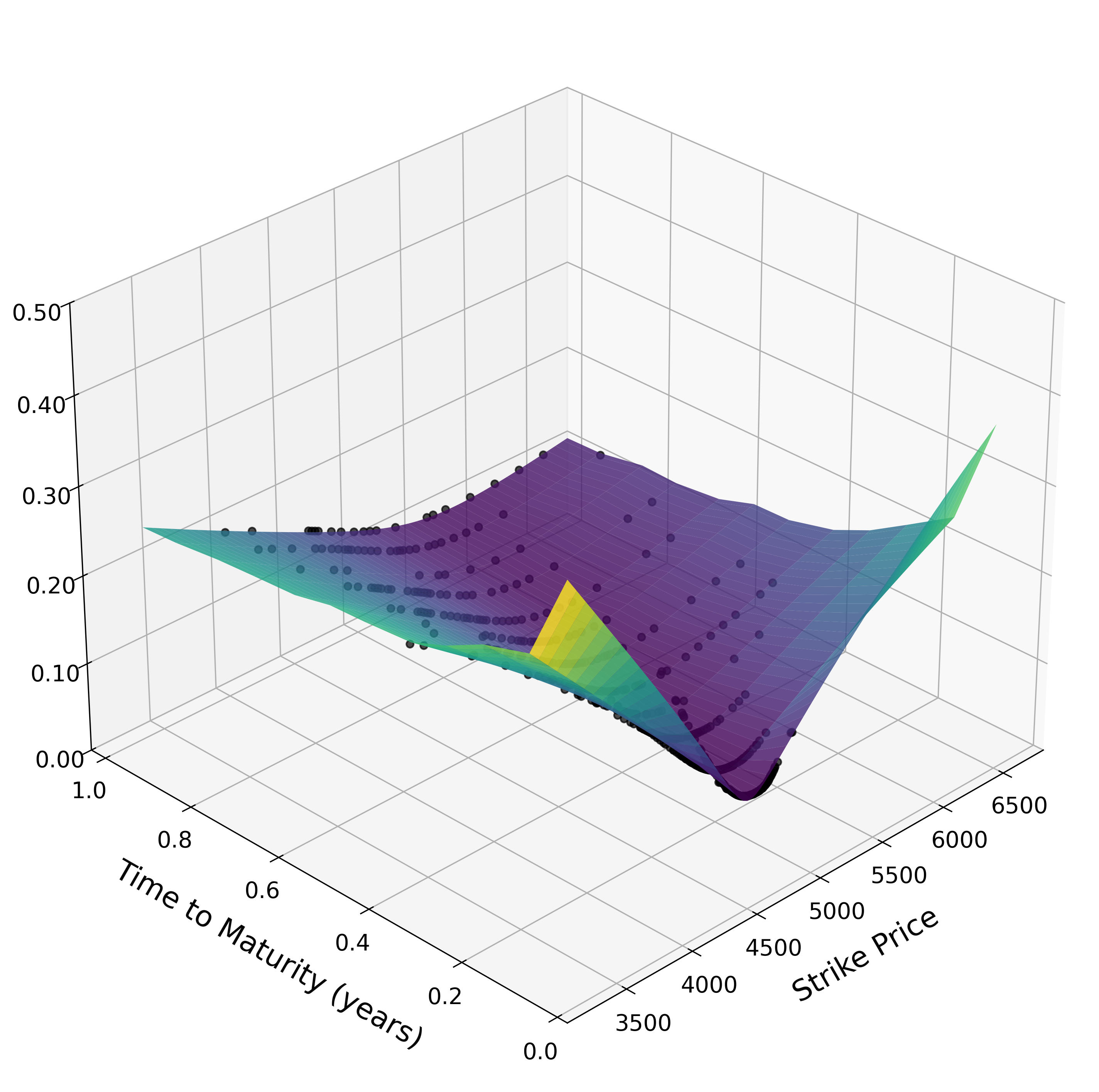}
        \caption{SABR}
        \label{fig:spx_surface_sabr_sub}
    \end{subfigure}
    \hfill
    \begin{subfigure}[b]{0.32\textwidth}
        \centering
        \includegraphics[width=\textwidth]{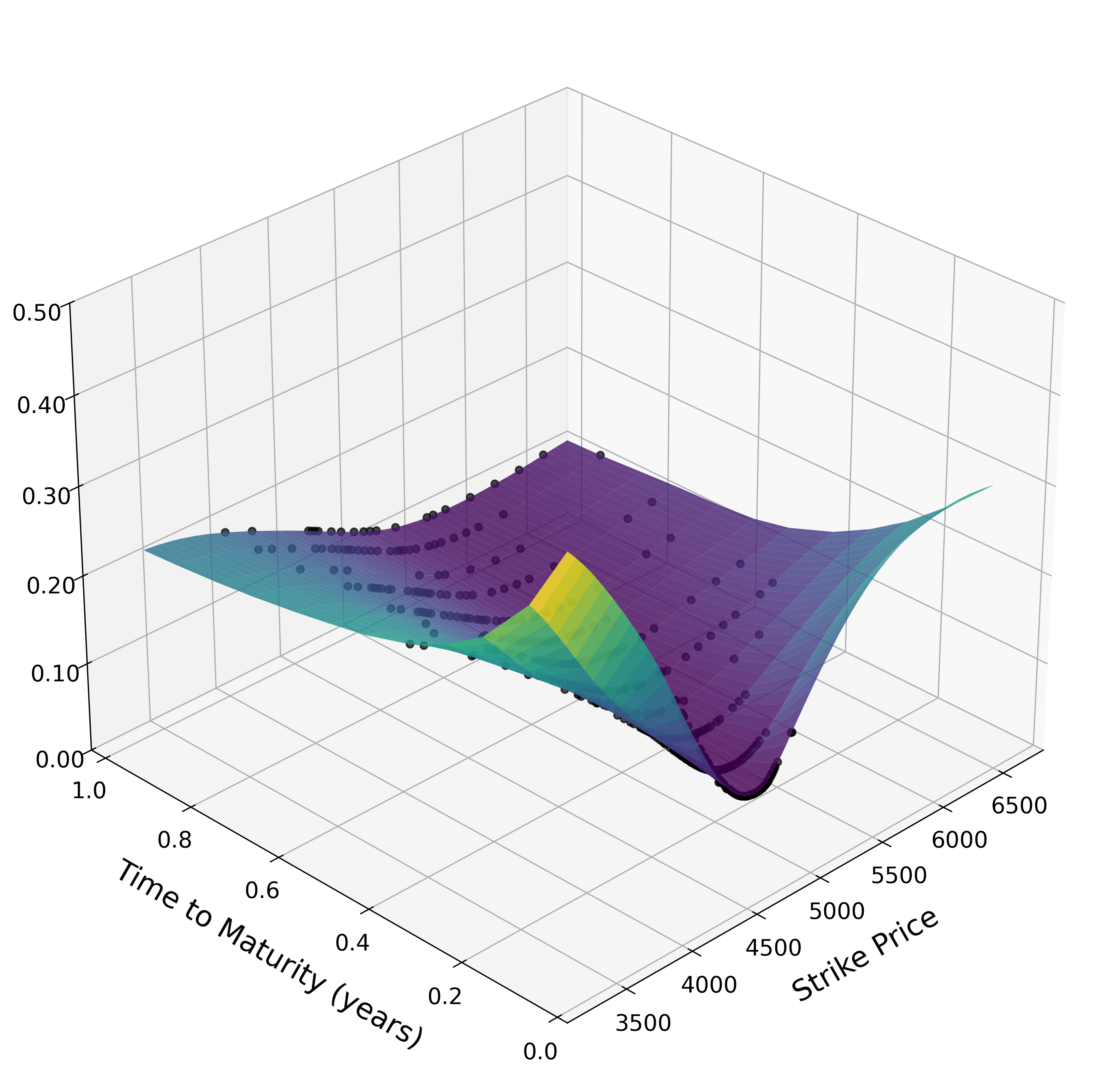}
        \caption{SABR-MTGP}
        \label{fig:spx_surface_mtgp_sub}
    \end{subfigure}
    \\
    \begin{subfigure}[b]{0.32\textwidth}
        \centering
        \includegraphics[width=\textwidth]{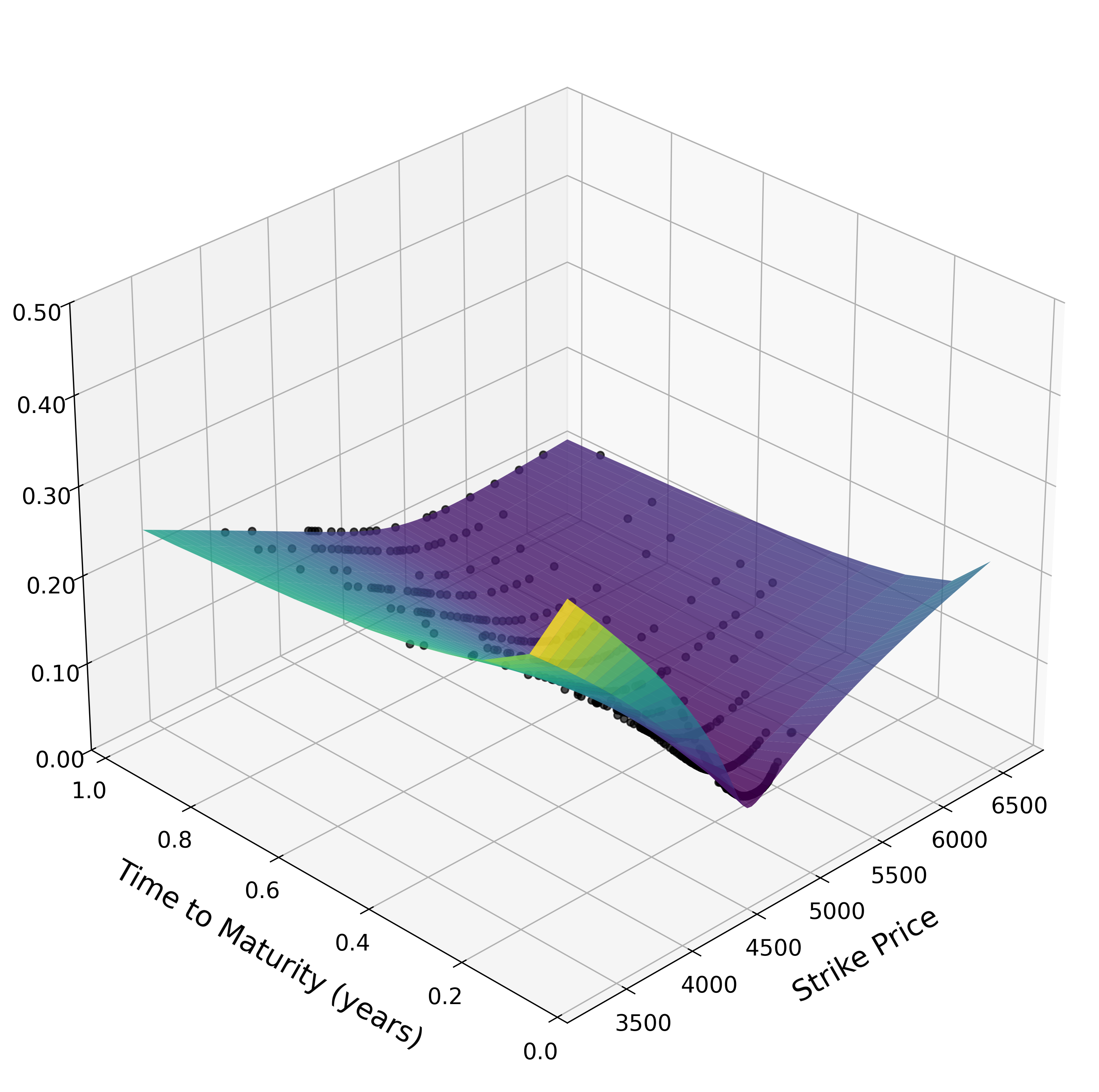}
        \caption{SSVI}
        \label{fig:spx_surface_ssvi_sub}
    \end{subfigure}
    \hfill
    \begin{subfigure}[b]{0.32\textwidth}
        \centering
        \includegraphics[width=\textwidth]{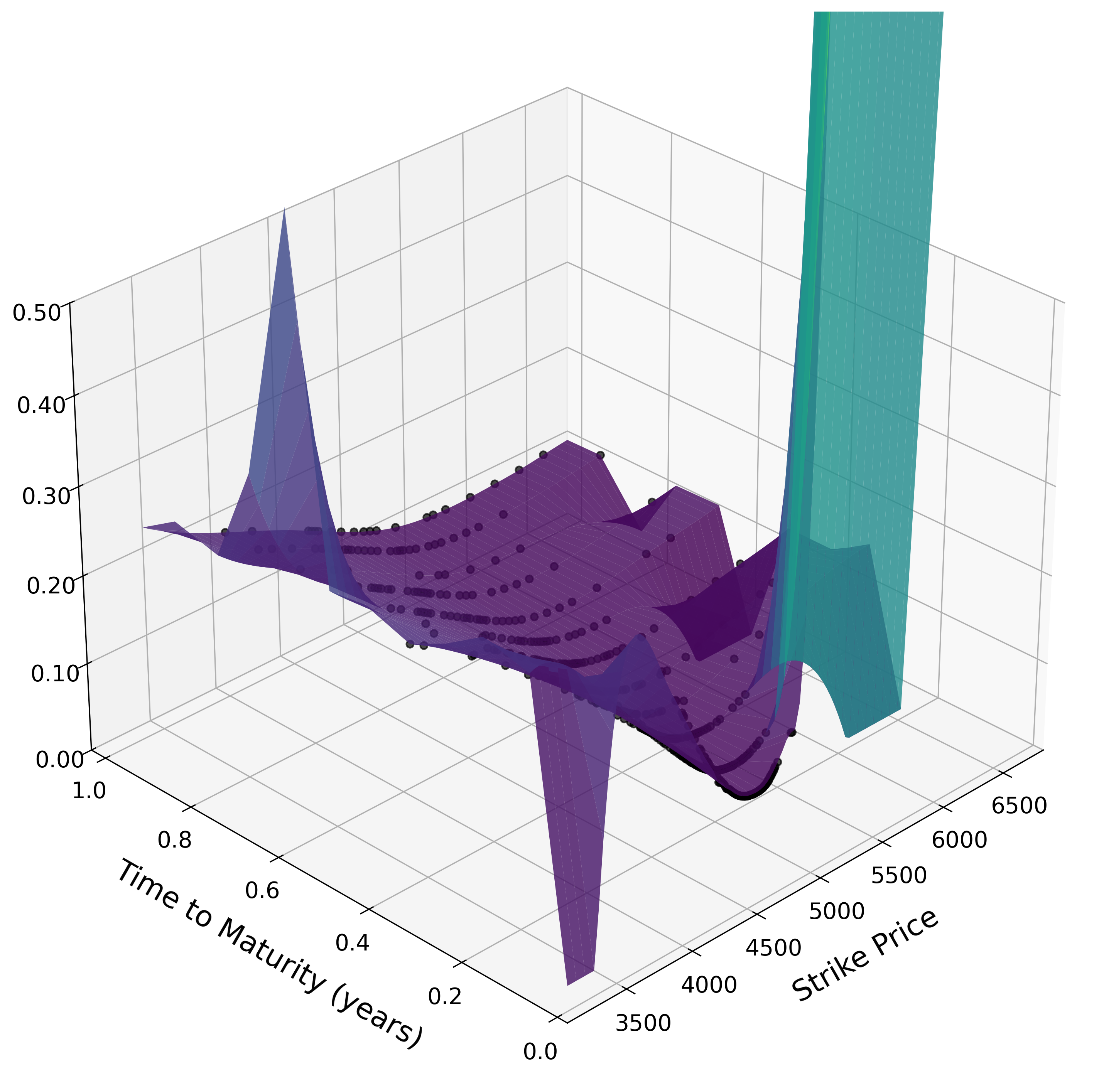}
        \caption{Cubic Spline}
        \label{fig:spx_surface_spline_sub}
    \end{subfigure}
    \hfill
    \begin{subfigure}[b]{0.32\textwidth}
        \centering
        \includegraphics[width=\textwidth]{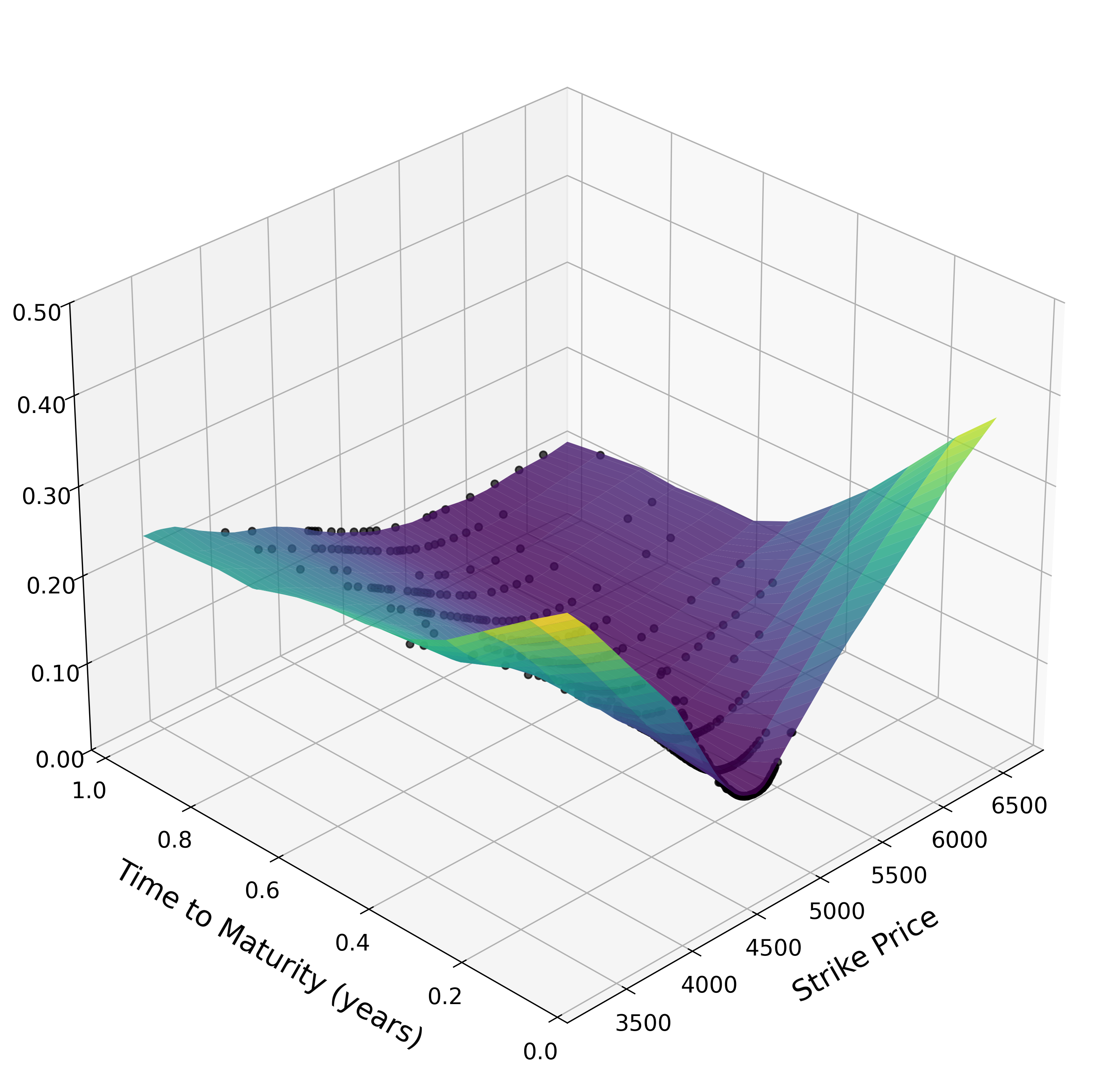}
        \caption{NN}
        \label{fig:spx_surface_nn_sub}
    \end{subfigure}
    \caption{SPX implied volatility surfaces constructed using six methods: GP, SABR, SABR-MTGP, SSVI, cubic spline, and a neural network. Black dots denote market observations.}
    \label{fig:spx_surfaces}
\end{figure}

Across the six surfaces in Figure~\ref{fig:spx_surfaces}, SABR-MTGP produces a smooth, well-behaved surface that tracks market levels across maturities and remains stable in sparse regions. The single-task GP is competitive overall but exhibits mild local undulations near the wings and toward the long end. The SABR and the neural network are smooth but can be too rigid around the wings. SSVI delivers a shape-constrained surface, whereas the cubic spline is generally smooth but shows occasional edge bending and a localized bulge at mid-maturities and ATM. Overall, the figures illustrate the intended trade-off: SABR-MTGP balances structural guidance and flexibility, preserving smoothness and plausibility while remaining close to the market points.

\begin{figure}
    \centering
    \includegraphics[width=0.8\linewidth]{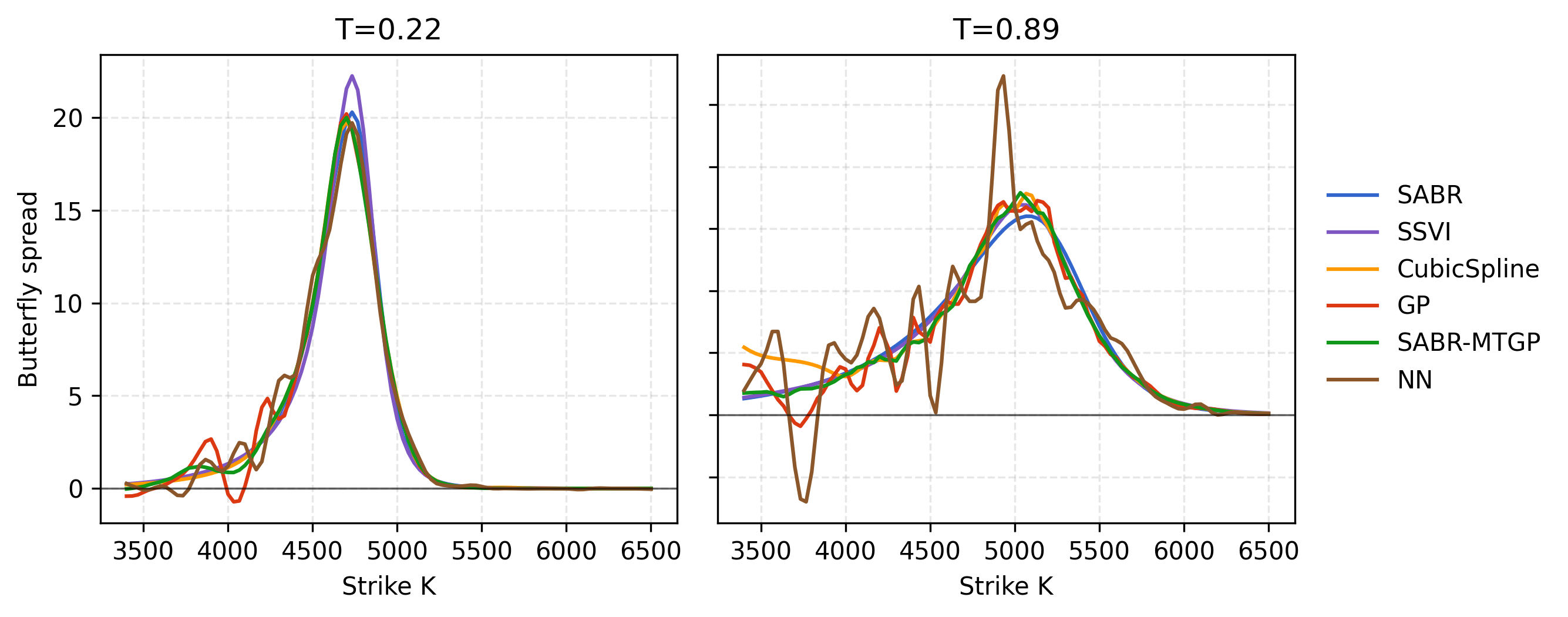}
    \caption{Butterfly spreads across strikes.}
    \label{fig:spx_butterfly}
\end{figure}

The butterfly spreads shown in Figure \ref{fig:spx_butterfly} remain nonnegative for SABR-MTGP, SABR, SSVI and cubic spline on the inspected slices, while GP and NN exhibit occasional dips near mid or long maturities. These diagnostics indicate that the surface generated by the proposed method is consistent with static no-arbitrage on the SPX sample.

\section{Conclusion}

In this paper, we proposed SABR-MTGP, an approach bridging structural and data-driven methods for IVS construction. The key idea is framing IVS construction as a multitask learning problem where we used structural information from the SABR model to improve predictions from sparse market data.

Our approach offered several contributions. First, we show how to combine financial models with machine learning techniques. We used SABR as a complementary information source within our multitask framework. Second, we developed a hierarchical Bayesian regularization for task embeddings that adaptively determined the level of information transfer. Third, our evaluation using Heston ground truth data showed this approach balanced structure and flexibility. It provided reliable predictions across market conditions and outperformed the standard Gaussian process, particularly in data-sparse regions. Improvements were competitive for longer maturities and extreme strikes, where market data was limited. Finally, an application to real SPX market data further illustrated the model's ability to produce a smooth, reliable IVS that respects market observations while benefiting from structural guidance.

Our work has practical implications for quantitative finance. SABR-MTGP provides a more reliable tool for IVS construction, crucial for risk management. Hedging and OTC valuation require market-consistent dynamic models (e.g., local or stochastic local volatility) and Greek validation. These topics are outside the scope of this paper. By incorporating structural knowledge and empirical observations, it reduced model misspecification risk while maintaining adaptability to market realities. Future research could extend the multitask framework to other structural models or multiple models simultaneously.
\bibliographystyle{elsarticle-num-names} 
\bibliography{reference}

@article{refWilliams2006,
  title={Gaussian processes for machine learning},
  author={Williams, Christopher KI and Rasmussen, Carl Edward},
  volume={2},
  number={3},
  year={2006},
  journal={MIT press Cambridge, MA}
}

@article{Hagan2002,
author = {Hagan, Patrick and Kumar, Deep and Lesniewski, Andrew and Woodward, Diana},
year = {2002},
month = {01},
pages = {84-108},
title = {Managing Smile Risk},
volume = {1},
journal = {Wilmott Magazine}
}

@article{bonilla2007multi,
  title={Multi-task Gaussian process prediction},
  author={Bonilla, Edwin V and Chai, Kian and Williams, Christopher},
  journal={Advances in neural information processing systems},
  volume={20},
  year={2007}
}

@article{alvarez2012kernels,
  title={Kernels for vector-valued functions: A review},
  author={Alvarez, Mauricio A and Rosasco, Lorenzo and Lawrence, Neil D and others},
  journal={Foundations and Trends in Machine Learning},
  volume={4},
  number={3},
  pages={195--266},
  year={2012},
  publisher={Now Publishers, Inc.}
}

@article{chen2023teaching,
  title={Teaching economics to the machines},
  author={Chen, Hui and Cheng, Yuhan and Liu, Yanchu and Tang, Ke},
  journal={Available at SSRN},
  volume={4642167},
  year={2023}
}

@article{heston1993closed,
  title={A closed-form solution for options with stochastic volatility with applications to bond and currency options},
  author={Heston, Steven L},
  journal={The review of financial studies},
  volume={6},
  number={2},
  pages={327--343},
  year={1993},
  publisher={Oxford University Press}
}

@article{black1973pricing,
  title={The pricing of options and corporate liabilities},
  author={Black, Fischer and Scholes, Myron},
  journal={Journal of political economy},
  volume={81},
  number={3},
  pages={637--654},
  year={1973},
  publisher={The University of Chicago Press}
}

@article{cousin2022gaussian,
  title={Gaussian process regression for swaption cube construction under no-arbitrage constraints},
  author={Cousin, Areski and Deleplace, Adrien and Misko, Adrien},
  journal={Risks},
  volume={10},
  number={12},
  pages={232},
  year={2022},
  publisher={MDPI}
}

@article{chataigner2021beyond,
  title={Beyond surrogate modeling: Learning the local volatility via shape constraints},
  author={Chataigner, Marc and Cousin, Areski and Cr{\'e}pey, St{\'e}phane and Dixon, Matthew and Gueye, Djibril},
  journal={SIAM Journal on Financial Mathematics},
  volume={12},
  number={3},
  pages={SC58--SC69},
  year={2021},
  publisher={SIAM}
}

@inproceedings{Ackerer2020,
author = {Ackerer, Damien and Tagasovska, Natasa and Vatter, Thibault},
title = {Deep smoothing of the implied volatility surface},
year = {2020},
isbn = {9781713829546},
publisher = {Curran Associates Inc.},
address = {Red Hook, NY, USA},
booktitle = {Proceedings of the 34th International Conference on Neural Information Processing Systems},
articleno = {969},
numpages = {12},
location = {Vancouver, BC, Canada},
series = {NIPS '20}
}

@inproceedings{zheng2021incorporating,
  title={Incorporating prior financial domain knowledge into neural networks for implied volatility surface prediction},
  author={Zheng, Yu and Yang, Yongxin and Chen, Bowei},
  booktitle={Proceedings of the 27th ACM SIGKDD Conference on Knowledge Discovery \& Data Mining},
  pages={3968--3975},
  year={2021}
}

@article{gonon2024operator,
  title={Operator Deep Smoothing for Implied Volatility},
  author={Gonon, Lukas and Jacquier, Antoine and Wiedemann, Ruben},
  journal={arXiv preprint arXiv:2406.11520},
  year={2024}
}

@article{hoshisashi2023no,
  title={No-Arbitrage Deep Calibration for Volatility Smile and Skewness},
  author={Hoshisashi, Kentaro and Phelan, Carolyn E and Barucca, Paolo},
  journal={arXiv preprint arXiv:2310.16703},
  year={2023}
}

@article{ Cousin2016,
Author = {Cousin, Areski and Maatouk, Hassan and Rulliere, Didier},
Title = {Kriging of financial term-structures},
Journal = {European Journal of Operational Research},
Year = {2016},
Volume = {255},
Number = {2},
Pages = {631-648},
Month = {12},
}

@article{Spiegeleer2018,
  title={Machine learning for quantitative finance: fast derivative pricing, hedging and fitting},
  author={De Spiegeleer, Jan and Madan, Dilip B and Reyners, Sofie and Schoutens, Wim},
  journal={Quantitative Finance},
  volume={18},
  number={10},
  pages={1635--1643},
  year={2018},
  publisher={Taylor \& Francis}
}

@article{risk2018sequential,
  title={Sequential design and spatial modeling for portfolio tail risk measurement},
  author={Risk, Jimmy and Ludkovski, Michael},
  journal={SIAM Journal on Financial Mathematics},
  volume={9},
  number={4},
  pages={1137--1174},
  year={2018},
  publisher={SIAM}
}

@article{ Ludkovski2018,
Author = {Ludkovski, Mike},
Title = {Kriging metamodels and experimental design for Bermudan option pricing},
Journal = {Journal of Computational Finance},
Year = {2018},
Volume = {22},
Number = {1},
Pages = {37-77},
Month = {01},

}

@article{ Crepey2020,
Author = {Crepey, Stephane and Dixon, Matthew F.},
Title = {Gaussian process regression for derivative portfolio modeling and
   application to credit valuation adjustment computations},
Journal = {Journal of Computational Finance},
Year = {2020},
Volume = {24},
Number = {1},
Pages = {47-81},
Month = {01},
}

@article{goudenege2020machine,
  title={Machine learning for pricing American options in high-dimensional Markovian and non-Markovian models},
  author={Goudenege, Ludovic and Molent, Andrea and Zanette, Antonino},
  journal={Quantitative Finance},
  volume={20},
  number={4},
  pages={573--591},
  year={2020},
  publisher={Taylor \& Francis}
}

@inproceedings{de2020gaussian,
  title={Gaussian process imputation of multiple financial series},
  author={de Wolff, Taco and Cuevas, Alejandro and Tobar, Felipe},
  booktitle={ICASSP 2020-2020 IEEE International Conference on Acoustics, Speech and Signal Processing (ICASSP)},
  pages={8444--8448},
  year={2020},
  organization={IEEE}
}

@article{roberts2021probabilistic,
  title={Probabilistic machine learning for local volatility},
  author={Roberts, Stephen and Tegn{\'e}r, Martin},
  journal={Journal of Computational Finance},
  volume={25},
  number={3},
  year={2021}
}

@article{goudenege2021gaussian,
  title={Gaussian process regression for pricing variable annuities with stochastic volatility and interest rate},
  author={Gouden{\`e}ge, Ludovic and Molent, Andrea and Zanette, Antonino},
  journal={Decisions in Economics and Finance},
  volume={44},
  number={1},
  pages={57--72},
  year={2021},
  publisher={Springer}
}

@article{ludkovski2021,
  title={KrigHedge: Gaussian process surrogates for delta hedging},
  author={Ludkovski, Mike and Saporito, Yuri},
  journal={Applied Mathematical Finance},
  volume={28},
  number={4},
  pages={330--360},
  year={2021},
  publisher={Taylor \& Francis}
}

@inproceedings{li2022effect,
  title={The Effect of Regression Methods on the Performance of Options Pricing using Machine Learning},
  author={Li, Lin and Gong, Zuoquan and Zheng, Li and Yu, Xi and Deng, Mingsen},
  booktitle={Proceedings of the 2022 13th International Conference on E-business, Management and Economics},
  pages={374--380},
  year={2022}
}

@article{xu2023gaussian,
  title={A Gaussian process regression machine learning model for forecasting retail property prices with Bayesian optimizations and cross-validation},
  author={Xu, Xiaojie and Zhang, Yun},
  journal={Decision Analytics Journal},
  volume={8},
  pages={100267},
  year={2023},
  publisher={Elsevier}
}

@article{hocht2024pricing,
  title={On the pricing of capped volatility swaps using machine learning techniques},
  author={H{\"o}cht, Stephan and Schoutens, Wim and Verschueren, Eva},
  journal={Quantitative Finance},
  volume={24},
  number={9},
  pages={1287--1300},
  year={2024},
  publisher={Taylor \& Francis}
}

@article{clark2024forecasting,
  title={Forecasting US inflation using Bayesian nonparametric models},
  author={Clark, Todd E and Huber, Florian and Koop, Gary and Marcellino, Massimiliano},
  journal={The Annals of Applied Statistics},
  volume={18},
  number={2},
  pages={1421--1444},
  year={2024},
  publisher={Institute of Mathematical Statistics}
}

@article{salvagnin2024investigating,
  title={Investigating the price determinants of the European Emission Trading System: a non-parametric approach},
  author={Salvagnin, Cristiano and Glielmo, Aldo and De Giuli, Maria Elena and Mira, Antonietta},
  journal={Quantitative Finance},
  volume={24},
  number={10},
  pages={1529--1544},
  year={2024},
  publisher={Taylor \& Francis}
}

@article{hauzenberger2025gaussian,
  title={Gaussian process vector autoregressions and macroeconomic uncertainty},
  author={Hauzenberger, Niko and Huber, Florian and Marcellino, Massimiliano and Petz, Nico},
  journal={Journal of Business \& Economic Statistics},
  volume={43},
  number={1},
  pages={27--43},
  year={2025},
  publisher={Taylor \& Francis}
}

@article{carr1999option,
  title={Option valuation using the fast Fourier transform},
  author={Carr, Peter and Madan, Dilip},
  journal={Journal of computational finance},
  volume={2},
  number={4},
  pages={61--73},
  year={1999}
}

@article{nelder1965simplex,
  title={A simplex method for function minimization},
  author={Nelder, John A and Mead, Roger},
  journal={The computer journal},
  volume={7},
  number={4},
  pages={308--313},
  year={1965},
  publisher={The British Computer Society}
}

@article{ning2023arbitrage,
  title={Arbitrage-free implied volatility surface generation with variational autoencoders},
  author={Ning, Brian and Jaimungal, Sebastian and Zhang, Xiaorong and Bergeron, Maxime},
  journal={SIAM Journal on Financial Mathematics},
  volume={14},
  number={4},
  pages={1004--1027},
  year={2023},
  publisher={SIAM}
}

@article{cont2023simulation,
  title={Simulation of arbitrage-free implied volatility surfaces},
  author={Cont, Rama and Vuleti{\'c}, Milena},
  journal={Applied Mathematical Finance},
  volume={30},
  number={2},
  pages={94--121},
  year={2023},
  publisher={Taylor \& Francis}
}

@article{corlay2013b,
  title={B-spline techniques for volatility modeling},
  author={Corlay, Sylvain},
  journal={arXiv preprint arXiv:1306.0995},
  year={2013}
}

@article{gatheral2004parsimonious,
  title={A parsimonious arbitrage-free implied volatility parameterization with application to the valuation of volatility derivatives},
  author={Gatheral, Jim},
  journal={Presentation at Global Derivatives \& Risk Management, Madrid},
  pages={0},
  year={2004}
}

@article{gatheral2014arbitrage,
  title={Arbitrage-free SVI volatility surfaces},
  author={Gatheral, Jim and Jacquier, Antoine},
  journal={Quantitative Finance},
  volume={14},
  number={1},
  pages={59--71},
  year={2014},
  publisher={Taylor \& Francis}
}

@article{madan1998variance,
  title={The variance gamma process and option pricing},
  author={Madan, Dilip B and Carr, Peter P and Chang, Eric C},
  journal={Review of Finance},
  volume={2},
  number={1},
  pages={79--105},
  year={1998},
  publisher={European Finance Association}
}

@article{kou2002jump,
  title={A jump-diffusion model for option pricing},
  author={Kou, Steven G},
  journal={Management science},
  volume={48},
  number={8},
  pages={1086--1101},
  year={2002},
  publisher={INFORMS}
}

@article{bayer2016pricing,
  title={Pricing under rough volatility},
  author={Bayer, Christian and Friz, Peter and Gatheral, Jim},
  journal={Quantitative Finance},
  volume={16},
  number={6},
  pages={887--904},
  year={2016},
  publisher={Taylor \& Francis}
}

@incollection{gatheral2022volatility,
  title={Volatility is rough},
  author={Gatheral, Jim and Jaisson, Thibault and Rosenbaum, Mathieu},
  booktitle={Commodities},
  pages={659--690},
  year={2022},
  publisher={Chapman and Hall/CRC}
}

@article{ WOS:000382423100006,
Author = {Zhang, Mengfei and Fabozzi, Frank J.},
Title = {On the Estimation of the SABR Model's Beta Parameter: The Role of
   Hedging in Determining the Beta Parameter},
Journal = {JOURNAL OF DERIVATIVES},
Year = {2016},
Volume = {24},
Number = {1},
Pages = {48-57},
Month = {FAL},
DOI = {10.3905/jod.2016.24.1.048},
ISSN = {1074-1240},
EISSN = {2168-8524},
Unique-ID = {WOS:000382423100006},
}

\end{document}